%% file: ttl.tex
\documentclass{sig-alternate}

\usepackage{amsmath}
\usepackage{amsfonts}
\usepackage{amssymb}
\usepackage{colordvi}
\usepackage{graphicx}
\usepackage{epsfig}
\usepackage{color}
\usepackage{multirow}
\usepackage{url}
\usepackage{bm}
\usepackage{caption}
\usepackage{tikz}
\usetikzlibrary{arrows,matrix,backgrounds,fit,calc,automata,shapes,positioning,decorations.pathreplacing}

\newcommand{\overbar}[1]{\mkern 1.5mu\overline{\mkern-2mu#1\mkern-6mu}\mkern 4.5mu}
\newcommand{\nooverbar}[1]{\mkern -0.2mu #1 \mkern -1mu}

\def\P{\mathbb{P}}

\def\et{\textit{et al.}}

\def\a{\textbf{a}}

\def\S{\textbf{S}}

\def\SR{$\min(\Sigma,\mathcal{R})$ }
\def\R{$\mathcal{R}$ }
\def\S{$\Sigma$ }
\def\SRN{$\min(\Sigma,\mathcal{R})$}
\def\RN{$\mathcal{R}$}
\def\SN{$\Sigma$}
\def\GGR{$G$-$G$-$\mathcal{R}$ }

\def\GMR{$G$-$M$-$\mathcal{R}$ }
\def\GMS{$G$-$M$-$\Sigma$ }
\def\MDR{$M$-$D$-$\mathcal{R}$ }
\def\MMR{$M$-$M$-$\mathcal{R}$ }
\def\GGS{$G$-$G$-$\Sigma$ }
\def\GGM{$G$-$G$-$\min(\Sigma,\mathcal{R})$ }
\def\MMmin{$M$-$M$-$\min(\Sigma,\mathcal{R})$ }

\def\MDS{$M$-$D$-$\Sigma$ }
\def\MMS{$M$-$M$-$\Sigma$ }
\def\GGmin{$G$-$G$-min$(\mathcal{R},\Sigma)$ }
\def\GGRN{$G$-$G$-$\mathcal{R}$}

\def\MMRN{$M$-$M$-$\mathcal{R}$}
\def\GGSN{$G$-$G$-$\Sigma$}

\def\MMSN{$M$-$M$-$\Sigma$}
\def\GGminN{$G$-$G$-min$(\mathcal{R},\Sigma)$}

\newcommand{\e}[1]{{\mathbb E }\left[ #1 \right]}
\newcommand{\ee}[1]{\mathbb{\tilde E}\left[ #1 \right]}

\newcommand{\pt}{\mathbb{\tilde P}}
\newcommand{\ft}{\tilde F}
\newcommand{\dft}{\tilde f}
\newcommand{\mgf}[1]{\phi( #1 )}

\newcommand\restr[2]{{
  \left.\kern-\nulldelimiterspace 
  #1 
  \vphantom{\big|} 
  \right|_{#2} 
  }}

\newtheorem{theorem}{Theorem}

\newtheorem{lemma}{Lemma}
\newtheorem{definition}{Definition}

\newtheorem{corollary}{Corollary}
\newtheorem{proposition}{Proposition}
\newtheorem{notation}{Notation}

 {\begin{list}{}%
         {\setlength{\leftmargin}{#1}}%
         \item[]%
 }
 {\end{list}}

\begin{document}

\title{Exact Analysis of TTL Cache Networks.\\The Case of Caching Policies driven by Stopping Times}

\numberofauthors{4}
\author{\alignauthor Daniel S. Berger \\\affaddr{University of Kaiserslautern} \hspace{-1em}%
\alignauthor Philipp Gland \\\affaddr{TU Berlin} %
\and
\alignauthor Sahil Singla \\\affaddr{Carnegie Mellon University} %
\alignauthor Florin Ciucu \\\affaddr{University of Warwick}
\vspace{-1em}
}


\maketitle \begin{abstract} TTL caching models have recently
regained significant research interest, largely due to their
ability to fit popular caching policies such as LRU. This paper
advances the state-of-the-art analysis of TTL-based cache networks
by developing two \textit{exact} methods with orthogonal
generality and computational complexity. The first method
generalizes existing results for line networks under renewal
requests to the broad class of caching policies whereby evictions
are driven by stopping times. The obtained results are further
generalized, using the second method, to feedforward networks with
Markov arrival processes (MAP) requests. MAPs are particularly
suitable for non-line networks because they are closed not only
under superposition and splitting, as known, but also under
input-output caching operations as proven herein for phase-type
TTL distributions. The crucial benefit of the two closure
properties is that they jointly enable the first exact analysis of
feedforward networks of TTL caches in great generality.
\end{abstract}



\section{Introduction}
Time-to-Live (TTL) caches decouple the eviction mechanisms amongst
objects by associating each object with a timer. When a timer expires
the corresponding object is evicted from the cache. This seemingly
simple scheme explicitly guarantees (weak) consistency, for which
reason it has been widely deployed in DNS and web caching. What has
recently however made TTL analytical models quite popular is a subtle
mapping between \textit{capacity-driven} (e.g., Least-Recently-Used
(LRU)) and \textit{TTL-based} caching policies. This mapping was
firstly established through a remarkably accurate approximation by
Che~\et~\cite{che2002hierarchical} for the popular LRU policy, which
was recently theoretically justified and extended to FIFO
(first-in-first-out) and RND (random eviction) policies
(Fricker~\et~\cite{fricker2012lru}), and further confirmed to hold for
broader arrival models (Bianchi~\et~\cite{bianchi2013check}), and even
in networks with several replication strategies
(Martina~\et~\cite{martina13}). Moreover,
Fofack~\et~\cite{fofack2012analysis} independently argued that TTL
caches capture the properties of LRU, FIFO, and RND policies, and,
remarkably, presented the first exact analysis for a line of TTL
caches. While the analysis of TTL caches is arguably simpler and more
general than the analysis of capacity-driven policies, the exact
analysis of TTL networks in general has remained an open problem.

When considering a cache network (including TTL-based), there are
two inherent network operations which complicate the analysis:
\textit{input-output} and \textit{superposition}. Given a caching
node serving a request (point) process for some object (i.e., the
\textit{input}), the corresponding miss process is a sample of the
request process at those points when the object is absent from the
cache. The exact characterization of the \textit{output} process
is challenging; for instance, the convenient and often assumed
memorylessness property of request processes would not be retained
by the corresponding miss process due to the TTL's inherent
filtering effect. The \textit{superposition} operation occurs when
merging miss processes from upstream caches into a new input
process. Since convenient statistical properties of the input
processes (e.g., the renewal property) are altered through
superposition, the analytical tractability of input-output
operations becomes even more complex.

In this paper we give the first exact analysis of caching
(feedforward) networks by jointly addressing broad classes of
request models, TTL distributions, and caching policies. The
request processes are either renewals or Markov arrival processes
(MAPs); the latter are dense in a suitable class of point
processes and generalize, in particular, the more popular
Markov-modulated Poisson processes. The TTLs follow general
distributions including phase-type (PH), which are dense within
the set of probability distributions on $[0,\infty)$. Moreover, we
consider an abstract model for TTL caching policies, whereby cache
evictions are driven by stopping times, and which captures in
particular three popular policies. The `$\mathcal{R}$' policy
regenerates the TTLs at every object's request and maps to the LRU
policy. The `$\Sigma$' policy regenerates the TTLs only at those
requests resulting in cache misses and maps to FIFO and RND
policies. Lastly, the `$\min(\Sigma,\mathcal{R})$' policy combines
the key features of `\R' and `$\Sigma$', i.e., weak consistency
guarantee and efficient utilization of cache space, respectively.

We structure our results in two parts. First we generalize the
recent results from Fofack~\et~\cite{fofack2012analysis} which cover
line networks, renewals requests, general TTL distributions, and
the `\RN' policy, by additionally covering the `$\Sigma$' and
`$\min(\Sigma,\mathcal{R})$' policies. The key analytical
contribution is a unified method to recursively characterize
input-output operations. This method is based on a suitable change
of measure technique using martingales to derive the Laplace
transform of a stopped sum, whereby the sum's stopping time
characterizes the caching policy. Leveraging certain martingale
properties to extend from deterministic to stopping times, we are
able to systematically analyze the three caching policies and
conceivably many others. The proposed method suffers however from
the same limitation as~\cite{fofack2012analysis}: since renewals are not
closed under superposition, unless Poisson, only lines of caches
can be (exactly) analyzed.

To address the annoying limitation of renewals' lack of
superposition closure, the second part of the paper advocates MAPs
to model request processes. The motivation to use MAPs is fairly
straightforward since MAPs are known to be closed under
superposition. What is remarkable, however, is that we are able to
prove that MAPs are also closed under the input-output operation
when the TTLs are described by PH distributions, for all the three
`$\mathcal{R}$', `$\Sigma$', and `\SRN' policies. In other words,
miss processes are also MAPs and trees of cache nodes can be
iteratively analyzed. As a side remark, MAPs are also closed under
a splitting operation, enabling thus the analysis of feedforward
networks.

The two proposed methods advance the state-of-the-art analysis of
TTL cache networks by providing the first exact results covering
broad request models, caching policies, and network topologies.
The second method in particular has the key feature of enabling
the first (exact) analysis of feedforward networks and is thus
conceivably more general than the first one. However, there is a
fundamental tradeoff between the two, which is driven by the state
explosion of MAPs under the superposition and input-output
operations. Therefore, while the first method is computationally
fast but restricted to line networks, the second has a much wider
applicability but suffers from a high computational complexity,
i.e., exponential in the number of caches. However, because the
underlying MAPs' matrices are sparse, the numerical complexity can
be significantly reduced by using appropriate numerical methods or
by formulating lower and upper bound stochastic models.

The rest of the paper is organized as follows. In
Section~\ref{sec:cmrw} we summarize related cache models and
discuss related work. In Section~\ref{sec:roadmap} we list model
definitions and key objectives for the analysis. In
Section~\ref{sec:lines} we present the change of measure technique
to address lines of caches with renewal requests. In
Section~\ref{sec:trees} we consider more general networks with MAP
requests. Finally, in Section~\ref{sec:concl}, we conclude the
paper. Some of the proofs and examples of MAP input-output
constructions are given in the Appendix.

\section{Cache Models and Related Work}\label{sec:cmrw}
Caching is implemented in many computer and communication systems,
such as CPUs, databases, or the Internet. Consequently, many
analytical models and techniques have been developed to study its
performance. Next we discuss related caching policies and results.

Caching policies can roughly be divided into two groups:
\textit{capacity-driven} and \textit{TTL-based} (see~Rizzo and
Vicisano~\cite{rizzo2000replacement}). In the former, objects'
evictions are driven by the arrivals of uncached objects and the
capacity constraint. In the latter, objects' evictions are
determined by individual timers. When compared, TTL-based cache
models are typically easier to analyze because the caching
behavior of different objects is decoupled and can be thus
represented in terms of independent point processes.

In this paper we address the following three TTL-based caching
policies, which differ in the behavior of the TTLs' resets and
eviction times:
\begin{enumerate}
\item \textit{Policy $\mathcal{R}$}: The TTL is reset with every
 request and an object is evicted upon the TTL's expiration.
\item \textit{Policy $\Sigma$}: The TTL is reset only at the times
of unsuccessful requests and an object is evicted upon the TTL's
expiration. \item \textit{Policy $\min(\Sigma,\mathcal{R})$}: Two
TTLs are reset in parallel according to the $\mathcal{R}$ and
$\Sigma$ policies, respectively, and an object is evicted upon the
expiration of \textit{either} of them.
\end{enumerate}

The $\mathcal{R}$ policy can be regarded as a TTL-based
correspondent of LRU caches~\cite{che2002hierarchical}.
$\mathcal{R}$ was properly formalized and analyzed by
Fofack~\et~\cite{fofack2012analysis}. In turn, the $\Sigma$ policy
can be regarded as a TTL-based correspondent of FIFO and RND (see,
e.g., Fricker~\et~\cite{fricker2012lru}); as a side remark, DNS
and web caching implement a variant of $\Sigma$.

Unlike $\mathcal{R}$ and $\Sigma$, the \SR policy has not yet been
formalized, although related implementations exist. For instance,
the default mechanism in Amazon ElastiCache to enforce an upper
bound on memory consumption is to implement an LRU policy on top
of evictions caused by TTLs~\cite{elasticache2013}. Another
example is the Squid web cache which explicitly uses both
$\mathcal{R}$ and $\Sigma$ TTLs; the former is set locally and
referred to as the ``Storage LRU Expiration
Age''~\cite{squid2013faq}, whereas the latter is set by content
owners.

\medskip

In the following we discuss related work starting with single
cache models. Early analytical models addressed capacity-driven
caches, which, contrary to their implementation simplicity, proved
to be difficult to analyze. Some of the classic works (e.g.,
King~\cite{king1971analysis} and
Gelenbe~\cite{gelenbe1973unified}) provided exact results for LRU,
FIFO, and RND policies. However, these results were argued to be
intractable by  Dan and Towsley~\cite{dan1990} and 
Jelenkovic~\cite{jelenkovic1999asymptotic}, who proposed instead
accurate and computationally fast approximations.

With respect to the second group of TTL based caches, the first analytical model for a
single $\Sigma$ cache under renewal arrivals and deterministic
TTLs was given by Jung~\et~\cite{jung03} who derived the
steady-state hit probability.  Under the same assumptions, Bahat
and Makowski~\cite{bahat2005measuring} extended this result to the
case of non-zero delays between the origin server and the cache,
and derived the hit probability for those requests which are
consistent with documents that undergo ongoing updates.  For a
single $\mathcal{R}$ cache, Fofack~\et~\cite{fofack2012analysis}
obtained the exact hit probability and other metrics for renewal
arrivals.

Next we revisit analytical results for cache networks.
Rosenzweig~\et~\cite{rosensweig2010approximate} applied the
approximation scheme from~\cite{dan1990} to networks of LRU caches
(at the expense however of errors up to 16\%).
Psaras~\et~\cite{psaras2011modelling} proposed a Markov chain
approximation for LRU caches which can then be linked together to
form tree networks by assuming each cache's miss process to be
Poisson. Under a similar approximation scheme (i.e., each cache's
request process is Poisson) Gallo~\et~\cite{gallo2012performance}
considered homogeneous tree networks under the RND policy. A
connection between the two domains of capacity-driven and
TTL-based policies was established by
Che~\et~\cite{che2002hierarchical} for a simple two-level LRU
cache network. With the strong case made by
Fricker~\et~\cite{fricker2012impact,fricker2012lru} on its wide
applicability, the so-called \emph{Che approximation} has recently
gained popularity: its impressive accuracy and generality was
confirmed by Bianchi~\et~\cite{bianchi2013check} and
Martina~\et~\cite{martina13}, and parallel extensions to several
other caching policies and replication strategies have also been
proposed.

The success of the Che approximation is due to a subtle mapping
from the domain of capacity-driven caches to the domain of
TTL-based caches. In the case of LRU, the key idea is to couple
the cache capacity with the durations that objects spent in a cache
under the condition that no further arrivals occur (i.e., \RN). By
assuming these (random) durations as deterministic and equal for every object, the
LRU model reduces to a TTL
model~\cite{che2002hierarchical,fricker2012lru} that is easier to
analyze.

To analyze TTL cache networks (e.g., as arising from the Che
approximation for a capacity-driven policy),
Martina~\et~\cite{martina13} rely on Poisson approximations of the
output processes (for both $\mathcal{R}$ and $\Sigma$ caches) and
report accurate results. Moreover,
Fofack~\et~\cite{fofack2012analysis} derived the first exact
results for a line of $\mathcal{R}$ caches and analyzed tree
networks by relying on a renewal approximation of superposed
processes.

Unlike these works, which assume an independent cache behavior,
another set of works considered hierarchies of caches where the
layers are synchronized by an aging mechanism: the TTL value at
child caches are set to coincide with the remaining TTL of parent
caches. In this way, Cohen and Kaplan~\cite{cohen2001aging} were
able to derive the miss rate for a two-level hierarchy, and
Cohen~\et~\cite{cohen2001performance} extended this result to
heterogeneous parent nodes. Remarkably, by ingeniously leveraging
the system's property that misses occur synchronously,
Hou~\et~\cite{hou2004expiration} were able to analyze trees of
caches under Poisson arrivals at the leaf caches.

We finally overview some recent related work on potential
applications. Borst~\et~\cite{borst2010distributed} addressed the
optimization of link utilization with a linear program for content
placement in a tree network. An important insight is that placing
caches close to the network edge often supersedes placing
additional caches within the network. This was further confirmed
through simulations by Psaras~\et~\cite{psaras2011modelling} and
Fayazbakhsh~\et~\cite{fayazbakhsh2013less} for LRU caches; in
particular, the latter argued that placing caches at the leaf
nodes of an ISP access tree does not sacrifice performance when
compared to omnipresent cache placement. A potentially interesting
application of our results is to analytically confirm whether this
insight holds in a much broader sense, e.g., for other network
topologies where different nodes can implement different caching
policies (i.e., either \RN, \SN, or \SRN).

\section{Roadmap}\label{sec:roadmap}
In this section we state the key objectives for analyzing lines of
caches and feedforward networks. First we give some general
definitions concerning some arbitrary node in a cache network.

\begin{definition}[Arrival/Input Process]\ \\
\label{def:amp} For each object, the arrival process is
represented as a point (counting) process $N(t)$. The
corresponding inter-arrival process is denoted by
$\{X_t\}_{t\geq1}$.
\end{definition}

When analyzing a cache network, one needs to characterize the
miss/output process relating two consecutive caches. We give the
definition in the renewal case.

\begin{definition}[Miss/Output Process]\ \\\label{def:imt}Let a caching node with
inter-request times and TTL's be given by two independent renewal
processes: $\{X_t\}_{t\geq1}$ and $\{T_t\}_{t\geq1}$. The
corresponding miss process is also a renewal process with the same
distribution as the stopped sum
\begin{equation}
  S_{\tau} := X_1 + \dots + X_{\tau}~,\label{eq:stau}
\end{equation}
where $\tau$ is a stopping time defined separately for each
caching (TTL) policy:
\begin{eqnarray}
  \textrm{Policy~}\mathcal{R}:~~&\tau&:=\min\{t:X_t>T_t\}\label{eq:str}\\
  \textrm{Policy~}\Sigma:~~&\tau&:=\min\{t:\sum_{s=1}^t\label{eq:sts}
  X_s>T_1\} \\
\nonumber  \textrm{Policy~}\min(\Sigma,\mathcal{R}):~~&\tau
  &:=\min\{\min\{t:\sum_{s=1}^tX_s>T_1^{\Sigma}\}, \\
  && \quad \quad \min\{t:X_t>T_t^{\mathcal{R}}\}\}~.\label{eq:stmin}
\end{eqnarray}
For the last policy, $T_1^{\Sigma}$ and $T_t^{\mathcal{R}}$ are
independent renewal processes.
\end{definition}
The corresponding definition for the non-renewal case can be
stated similarly and is omitted for brevity.

As a side remark, the structure of the first two stopping times
justifies the notation for the caching policies by $\mathcal{R}$
and $\Sigma$: the former has a renewal structure (whence the
letter $\mathcal{R}$), whereas the latter has a sum structure
(whence the letter $\Sigma$). We incorporate this notation in the
whole notation of a caching node, by borrowing from Kendall's
notation in queueing theory.
\begin{notation}[Caching Node]\ \\Depending on the TTL policy, a caching node is denoted as
either one of the triplets
\begin{equation*}
\textrm{\GGRN~or~\GGSN~or~\GGminN}~,
\end{equation*}
where the two $G$'s stand for the generic distributions of the
inter-arrival times and the TTLs, respectively.
\end{notation}
As an example, a cache with exponentially distributed
inter-arrival and TTL times, and implementing the $\mathcal{R}$
policy, is denoted by \MMRN. Some other distributions used in this
paper are the deterministic (D) case, the exponential (M), and the
phase-type (PH) distribution.

\medskip

Cache performance is commonly measured in terms of hit/miss
probabilities, which indicate the improvement in link utilization
when using a cache. For example, the cost metrics
in~\cite{fayazbakhsh2013less} depend on the hit/miss probabilities
and the arrival rates at the leaves of a caching tree.

\begin{definition}[Hit/Miss Probability]\ \\
\label{def:hmp} Consider an arbitrary cache with arrival process
$N(t)$ and miss process $M(t)$, for some fixed object. The hit and
miss probabilities are defined as
\begin{eqnarray*}
&&H := \lim_{t \rightarrow \infty} \left(1 -
  \frac{M(t)}{N(t)}\right )\\
&&M := \lim_{t \rightarrow \infty} \frac{M(t)}{N(t)}~,
\end{eqnarray*}
respectively, subject to convergence.
\end{definition}

Another metric of interest is the cache occupancy which defines the
average amount of storage required by an object, and also establishes
a connection between capacity-driven and TTL-based cache models
through a suitable set of parameters. For example, the connection
between an LRU cache's capacity and the corresponding
TTL-$\mathcal{R}$ model is established by equalizing the summation of
the occupancies of the objects in the TTL cache with the LRU
capacity~\cite{che2002hierarchical}.

\begin{definition}[Cache Occupancy]\ \\
\label{def:co} Let $C(t)$ be a random binary process representing
whether the object is in the cache or not at time $t$. The cache
occupancy is defined as
\begin{eqnarray*}
\pi:=\lim_{t\rightarrow\infty}\frac{\int_0^t C(s)ds}{t}~.
\end{eqnarray*}
\end{definition}

\medskip

We next briefly introduce the key objectives for analyzing lines
and feedforward networks.

\subsection{Lines of Caches with Renewal Arrivals}

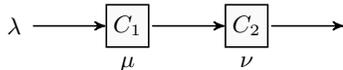
\begin{figure}[ht]
\begin{center}\footnotesize
\begin{tikzpicture}[<-,>=stealth',shorten >=1pt,auto,node distance=1cm,
                    semithick]
  \tikzstyle{every state}=[rectangle,fill=black!2,minimum size=16pt,inner sep=0pt]

\node[draw=none]                               (1) {$$};
\node[state,left=of 1]                    (2) {$C_2$};
\node[below=of 2,yshift=1cm]                    (2b) {$\nu$};
\node[state,left=of 2] (4) {$C_1$}; \node[below=of 4,yshift=1cm]
(2b) {$\mu$}; \node[draw=none,left=of 4](5) {$\lambda$}; \draw
     (1)   edge node {}   (2)
     (2)   edge node {}   (4)
     (4)   edge node {}   (5);
\end{tikzpicture}
\caption{A line of two caches $C_1$ and $C_2$ with Poisson
arrivals with rate $\lambda$ and exponential TTLs with rates
$\mu~\textrm{and}~\nu$}\label{fig:renewalcache}
\vspace{-2em}
\end{center}
\end{figure}

Consider the simplified line network with two nodes from
Figure~\ref{fig:renewalcache}. At the first node, requests for
some object arrive according to a renewal process
$\{X_t\}_{t\geq1}$. If the object is in the cache at the time of a
request, then the request is successful and the object is fetched.
Otherwise, for every unsuccessful request at some node, the object
is recursively requested at the next node in the line. Once the
object is successfully found at some downstream node, it is
(instantaneously) transferred to the upstream nodes. For the
model's completeness we assume that the last node always has a
copy of the object. All the nodes implement one of the
$\mathcal{R}$, $\Sigma$, or \SR caching policies (different nodes
are allowed to implement different policies).

We address a single node with a given arrival/input process and
TTL distribution. For this setting our objective is to derive the
Laplace transform of the corresponding miss/output process, i.e.,
of the stopped random walk $S_{\tau}$ from Eq.~(\ref{eq:stau}):
\begin{equation}
\mathcal{L}_{\omega}(S_\tau):=E\left[e^{-\omega\left(X_1+X_2+\dots+X_{\tau}\right)}\right]~.\label{eq:mgfstau}
\end{equation}
This technique can be iteratively applied along an entire line of
caches using numerical methods (as in
Fofack~\et~\cite{fofack2012analysis}); note that numerical methods are
not necessary in the case of exponential TTLs. We emphasize that,
unlike \cite{fofack2012analysis} which is restricted to $\mathcal{R}$,
our method additionally covers $\Sigma$ and
$\min(\Sigma,\mathcal{R})$.

\subsection{Feedforward Networks with MAP Arrivals} The
previous technique suffers from the major limitation that cache
requests must be a renewal process and thus it does not apply to
more general topologies subject to a superposition operation.

\begin{minipage}{0.15\textwidth}
\begin{tikzpicture}[<-,>=stealth',shorten >=1pt,auto,node distance=0.4cm,
                    semithick]
  \tikzstyle{every state}=[rectangle,fill=black!2,minimum size=16pt,inner sep=0pt]
\node[draw=none]                           (1)   {};
\node[state,below=of 1]                    (2)   {$C_3$};
\node[right=of 2, xshift=-0.45cm]           (2m)  {$\mu_3$};
\node[draw=none,below=of 2, scale=0.01]    (3)   {};
\node[state,below=of 3, xshift=-0.6cm]     (12)  {$C_1$};
\node[right=of 12, xshift=-0.45cm]          (12m) {$\mu_1$};
\node[draw=none,below=of 12]               (11)  {$\lambda_1$};
\node[state,below=of 3, xshift=0.6cm]      (22)  {$C_2$};
\node[right=of 22, xshift=-0.45cm]          (22m) {$\mu_2$};
\node[draw=none,below=of 22]               (21)  {$\lambda_2$};
\draw
     (1)   edge node {}   (2)
     (2)   edge node {}   (3)
     (12)   edge node {}   (11)
     (22)   edge node {}   (21);
\draw[-]
     (3)   edge node {}   (12)
     (3)   edge node {}   (22);
\end{tikzpicture}

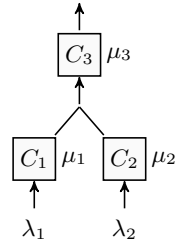
\captionof{figure}{A tree\\ of caches} \label{fig:treecaches}
\end{minipage}
\begin{minipage}{0.30\textwidth}
Indeed, consider the simple tree topology from
Figure~\ref{fig:treecaches} in which the inter-request times at
the leaf caches are $\sim\exp(\lambda_1)$ (exponential) and
$\sim\exp(\lambda_2)$, and the corresponding TTLs are
$\sim\exp(\mu_1)$ and $\sim\exp(\mu_2)$, respectively; all the
processes are independent. The inter-miss times at the leaf caches
are renewal processes with $hypo(\lambda_1,\mu_1)$
(hypoexponential) and $hypo(\lambda_2,\mu_2)$ distributions (to be
shown in Table~\ref{tb:examples}). The superposition of the two
renewal processes is not a renewal process, which means that the
\end{minipage}
technique targeting lines of caches does not apply at the root
cache (the superposition of renewal processes is a renewal process
if and only if the superposed processes are Poisson).

Let us now recall the two main analytical operations which must be
accounted for when analyzing trees of caches:
\begin{enumerate}
\item\textit{input-output}: the characterization of the inter-miss
process from the inter-request process.
\item\textit{superposition}: the characterization of a single
inter-request process from multiple ones (e.g., at the root cache
from Figure~\ref{fig:treecaches}).
\end{enumerate}

Unlike the \textit{input-output} operation which will be shown in
Section~\ref{sec:lines} to be tractable (yet subject to recursions and
also evaluation of convolutions in the case of $\Sigma$ and \SR
caches), the \textit{superposition} operation is conceivably the
bottleneck due to the lack of closure for renewal processes.  To
circumvent this apparent difficulty, the natural generalization of
renewal processes are Markov arrival process (MAPs), which are known
to be closed under \textit{superposition} (and also splitting). The
remaining objective is to additionally show that MAPs are also closed
under the \textit{input-output} operation of caches (see
Section~\ref{sec:trees}).

We point out that the apparently straightforward idea of using
MAPs has been efficiently used in the past to model systems with
non-renewal behavior, e.g., single queues with non-renewal
arrivals (Lucantoni~\et~\cite{lucantoni1990single}) or closed
queueing networks with non-renewal workloads
(Casale~\et~\cite{Casale08}); for an excellent related survey see
Asmussen~\cite{Asmussen00}.

\section{Lines of Caches}\label{sec:lines}
In this section we propose a unified method to analyze lines of
$G$-$G$-$\mathcal{R}$, $G$-$G$-$\Sigma$, and \GGM caches. First we
instantiate the caching metrics from Definitions~\ref{def:hmp} and
\ref{def:co} for the renewal case.

\begin{lemma}[Hit/Miss Probabilities]\label{lm:hmp}\ \\
For \GGRN, \GGSN, and \GGmin caches, the hit and miss
probabilities from Definition~\ref{def:hmp} become
\begin{align}
H = \frac{\e{\tau} - 1}{\e{\tau}} \text{ and } M =
\frac{1}{\e{\tau}}~.\label{eq:hmprob}
\end{align}
In particular, for \GGRN, it holds
\begin{align}
H = \P\left(X\leq T\right) \text{ and } M =\P\left(X>T\right)~.
\end{align}
\end{lemma}

The expression of the miss probability for the \GGS cache is the
same as in Jung~\et~\cite{jung03}. Unlike in the \GGR case,
$E[\tau]$ for the other two policies cannot be generally given in
closed-form due to the underlying convolution in the definition of
$\tau$ from Eqs.~(\ref{eq:sts})-(\ref{eq:stmin}); it is, however,
often straightforward to derive $E[\tau]$ for particular
distributions of $\{X_t\}_{t\geq1}$ and $\{T_t\}_{t\geq1}$.

\proof Let $t_n$ denote the point process of the unsuccessful request
times (i.e., the miss times) for $n\geq1$ and $t_0=0$. Using the
renewal property of $\{X_t\}_{t\geq1}$ and $\{T_t\}_{t\geq1}$, and
the strong law of large numbers, we have
\begin{equation*}
\lim_{n\rightarrow\infty}\frac{M(t_n)}{N(t_n)}
=\lim_{n\rightarrow\infty}\frac{n}{\tau_1+\dots+\tau_{n}}
=\frac{1}{E[\tau]}~,
\end{equation*}
where $\tau_i$ denotes the stationary sequence of stopping times,
as defined in Eqs.~(\ref{eq:str}) and (\ref{eq:sts}), but starting
from $t\geq t_{i-1}$ in the usual renewal sense. Moreover, since
for $t\in(t_{i-1},t_i]$
\begin{equation*}
\frac{M(t_{i})-1}{N(t_{i})}<\frac{M(t)}{N(t)}\leq\frac{M(t_{i-1})+1}{N(t_{i-1})}~,
\end{equation*}
the limit $\lim_{t\rightarrow\infty}\frac{M(t)}{N(t)}$ exists and
Eq.~(\ref{eq:hmprob}) is proven. The particular expression for
\GGR cache follows directly from the geometric distribution of
$\tau$.\hfill $\Box$

\begin{lemma}[Cache Occupancy]\label{lm:co}\ \\
For the \GGRN, \GGSN, and \GGM caches, the cache occupancies from
Definition~\ref{def:co} become
\begin{eqnarray*}
\pi_{\mathcal{R}} &=& \frac{\e{min\{X,T\}}}{\e{X}},~\pi_\Sigma =
\frac{\e{T}}{\e{S_\tau}}\label{eq:coren}\\
\pi_{\min(\Sigma,\mathcal{R})}&=&\frac{E\left[\min\{\sum_{s=1}^{\tau}\min\{X_s,T_s^{\mathcal{R}}\},T_1^{\Sigma}\}\right]}{E\left[S_{\tau}\right]}~.
\end{eqnarray*}
\end{lemma}

The expression for \GGR is the same as the one given by
Fofack~\et~\cite{fofack2012analysis} (written therein in the
equivalent form $\pi_{\mathcal{R}} =
\e{\int_0^X\P(T>t)dt}/\e{X}$). Note that the last expectation
depends on the stopping time $\tau$ from Eq.~(\ref{eq:stmin}),
whose mass function is later provided in Corollary~\ref{lm:ggmin};
moreover, to compute the cache occupancy, a decoupling argument
like the one we provide for the transforms of the inter-miss
times is needed to avoid the implicit correlations amongst the
stopping time, the inter-arrival times, and the TTLs (for all
$\mathcal{R}$, $\Sigma$, and \SRN).

\proof In the case of \GGRN, denote the point process $t_n$ of the
request times, i.e., $t_n=\sum_{i=1}^n X_i$. Using the renewal
property of $\{X_t\}_{t\geq1}$ and $\{T_t\}_{t\geq1}$, and the
strong law of large numbers, we have
\begin{eqnarray*}
\lim_{n\rightarrow\infty}\frac{\int_0^{t_n}C(s)ds}{t_n}&=&\lim_{n\rightarrow\infty}\frac{\sum_{i=1}^n\min\{X_i,T_i\}}{\sum_{i=1}^n X_i}\\
&=&\frac{E\left[\min\{X,T\}\right]}{E[X]}~.
\end{eqnarray*}

In the case of \GGSN, we use the same embedding $t_n$ as in the
proof of Lemma~\ref{lm:hmp} such that
\begin{eqnarray*}
\lim_{n\rightarrow\infty}\frac{\int_0^{t_n}C(s)ds}{t_n}&=&\lim_{n\rightarrow\infty}\frac{\sum_{i=1}^n T_i}{\sum_{i=1}^n S_{\tau_i}}\\
&=&\frac{E\left[T\right]}{E\left[S_{\tau}\right]}~,
\end{eqnarray*}
where
$S_{\tau_i}:=X_{\tau_{i-1}+1}+\dots+X_{\tau_{i-1}+\tau_{i}}$.

In both cases, the extensions of the limits to the whole line
follows by a bounding argument as in the proof of
Lemma~\ref{lm:hmp}. The proof for \GGM follows using the same
embedding points as for \GGSN.\hfill $\Box$

\medskip

The key problem to obtain caching metrics at the downstream nodes
in a line of cache is to recursively characterize the inter-miss
time $S_{\tau}$ defined in Eq.~(\ref{eq:stau}). The expression of
$S_{\tau}$, as well as those of the above cache metrics, suggests
following a martingale based technique to characterize $S_{\tau}$.
This (rough) idea is driven by the fact that stopping
times---which are at the core of the very definition of
$S_{\tau}$---preserve certain martingale results, e.g., if $L_t$
is a martingale and $\tau$ is a bounded stopping time then
$E\left[L_{\tau}\right]=E\left[L_1\right]$, which is a particular
case of the optional stopping theorem. Note however that caching
stopping times are not necessarily bounded, and thus require a
more general framework. For example, the stopping times from
Eqs.~(\ref{eq:str})-(\ref{eq:stmin}) are under realistic
assumptions almost surely finite but may be unbounded.

Next we will demonstrate the effectiveness of relying on
martingale techniques to derive an elegant and unified analysis of
\GGRN, \GGSN, and \GGM caches. To this end, we first provide a
closed-form result (in most of the cases) for the Laplace
transform of the stopped random sum $S_{\tau}$. This result will
be instrumental to the analysis of the $\mathcal{R}$, $\Sigma$,
and \SR policies.

\subsection{The Laplace Transform of a Stopped Sum}\label{sec:mgfsrw}
Consider the two independent renewal processes $\{X_t\}_{t\geq1}$ and
$\{T_t\}_{t\geq1}$ on a joint probability space
$(\Omega,\mathcal{F},\mathbb{\P})$ (e.g., as in
Definition~\ref{def:imt}).  Denote the corresponding distribution
functions by $F(x)$ and $G(x)$, and assume the existence of
corresponding densities $f(x)$ and $g(x)$, respectively. Let
$\mathcal{F}_t=\sigma((X_1,T_1), \dots, (X_t,T_t))$,
$\mathbb{F}=(\mathcal{F},\{\mathcal{F}_t\}_{t\geq1})$ denote the
filtration associated with $S_\tau$, and let
$(\Omega,\mathbb{F},\mathbb{P})$ denote the corresponding filtered
probability space. When clear from the context the time indexes are
suppressed.

Next we provide a closed-form expression for the Laplace transform of
the stopped random walk $S_{\tau}$ from Eq.~(\ref{eq:mgfstau}).
Recall from Eqs.~(\ref{eq:str})-(\ref{eq:stmin}) that $\tau$ is a
stopping time with respect to the filtration $\mathcal{F}_t$. One
may remark that, due to the intrinsic dependencies amongst $X_t$'s
and the stopping time $\tau$, the analysis of the stopped sum
$S_{\tau}$ is conceivably quite involved even for stopping times
w.r.t. the filtration $\mathcal{F}_t'=\sigma(X_1,X_2,\dots,X_t)$. In
fact, unlike the first moment which is relatively easily obtained
as Wald's equation, i.e., $E\left[S_{\tau}\right]=E[\tau]E[X_1]$
(under the additional condition that $E[\tau]<\infty$), higher
moments, however, are typically only known in terms of bounds (see
Gut~\cite{Gut09},~p. 22).

Despite the apparent technical difficulties, we will next show
that $\mathcal{L}_{\omega}(S_\tau)$, for stopping times $\tau$ w.r.t.
$\mathcal{F}_t$, can be derived in a rather straightforward
manner. The key idea is to construct a suitable new filtered
probability space $(\Omega,\mathbb{F},\tilde{\mathbb{P}})$,
whereby the new probability measure $\tilde{\mathbb{P}}$ decouples
the dependencies amongst $X_t$'s and $\tau$. Informally, the key
idea to obtain $\mathcal{L}_{\omega}(S_\tau)$ in closed-form is to
\textit{offshore} the underlying derivations into the new
$(\Omega,\mathbb{F},\tilde{\mathbb{P}})$ space.

This technique is known as \textit{change of measure}. The change
of measure itself (e.g., from $\mathbb{P}$ to
$\tilde{\mathbb{P}}$) is performed as such \textit{measures} in
the original space (e.g., $\mathbb{P}(A)$ for $A\in{\mathcal{F}}$)
can be obtained in terms of the new (changed) measure in a much
simpler manner. An example of an application of this technique is
in rare events simulations, whereby rare events become more likely
to occur under the new (changed) measure, or more precisely under
the new (changed) density, guaranteeing thus faster convergence
speeds than Monte-Carlo simulations (see
Pham~\cite{pham2010largedev}). Another application is in pricing
risks in incomplete markets, by constructing a new risk-neutral
probability measure (see Cox~\et~\cite{cox2006multivariate}). Such
risk-neutral measures have also been constructed in financial
models, in order to simplify a model with drift into one with
constant expectation and allowing thus the application of the
Girsanov theorem to describe the process dynamics (see Musiela and
Rutkowski~\cite{musiela2005martingale}). Another application is an
elegant proof for Cram\'{e}r's theorem in large deviation theory
(see Dembo and Zeitouni~\cite{Dembo98}, p. 27).

To perform the intended change of measure, we extend the measure
construction for a filtration
$\mathcal{F}_t'=\sigma(X_1,X_2,\dots,X_t)$ (see
Asmussen~\cite{Asmussen03}, p. 358) to the product filtration
$\mathcal{F}_t=\sigma((X_1,T_1),\dots,(X_t,T_t))$ which accounts
for $(T_t)_{t\geq1}$ as well. While the extension proceeds mutatis
mutandis, mainly due to the independence between $(X_t)_{t\geq1}$
and $(T_t)_{t\geq1}$, the key to our construction is to only
\textit{tilt} the distribution $F(x)$ of $X_t$ while preserving
the distribution $G(x)$ of $T_t$; for this reason, we refer to our
change of measure as a \textit{fractional} change of measure.

\begin{definition}[Fractional Change of Measure]\
\\\label{def:fcm}
For any $F\in\mathcal{F}_t$ define the tilted probability measure
$\tilde{\mathbb{P}}_t$ as
\begin{equation}
\tilde{\mathbb{P}}_t(F):= E\left[L_t 1_F\right]~,
\end{equation}
where $L_t$ is the Wald's martingale
\begin{equation}
L_t:=\frac{e^{-\omega
S_t}}{\mathcal{L}_{\omega}(X)^t}~\forall~t\geq1\label{eq:waldmart}
\end{equation}
w.r.t. the original filtered space
$(\Omega,(\mathcal{F},\{\mathcal{F}_t'\}_{t\geq1}),\mathbb{P})$,
and for some fixed $\omega\in\mathbb{R}$.
\end{definition}

The tilted measures $\tilde{\mathbb{P}}_t$, which are by
construction restricted to $\mathcal{F}_t$, uniquely extend to a
probability measure $\tilde{\mathbb{P}}$ on $\mathcal{F}$ which is
Kolmogorov consistent, i.e.,
\begin{equation*}
\tilde{\mathbb{P}}(F) = \tilde{\mathbb{P}}_t(F)=E\left[L_t
1_F\right]~,
\end{equation*}
for all $F\in\mathcal{F}_t$. The proof follows the proof of
Proposition~3.1 from~\cite{Asmussen03}, with the observation that
$L_t$ is also a martingale w.r.t. the product filtered space
$(\Omega,\mathbb{F},\mathbb{P})$ due to the independence between
$(X_t)_{t\geq1}$ and $(T_t)_{t\geq1}$.

Besides enabling the construction of the consistent probability
measure $\tilde{\mathbb{P}}$ (mainly using the fact that $L_t$'s
are martingales with $E\left[L_t\right]=1$), there are two
technical reasons behind the fractional change of measure from
Definition~\ref{def:fcm}. On one hand, $L_t$ corresponds to the
Radon-Nikodym density of the Kolmogorov extended measure
$\tilde{\mathbb{P}}$ (in addition to that of $\mathbb{P}_t$ as
well) w.r.t. $\mathbb{P}$, i.e.,
$L_t=\frac{d\tilde{\mathbb{P}}}{d\mathbb{P}}$ on
$(\Omega,\mathcal{F}_t)$ for all $t\geq1$. This allows the
computation of integrals w.r.t. $\tilde{\mathbb{P}}$ according to the
integration rule
\begin{equation*}
\int_{A}Yd\tilde{\mathbb{P}}=\int_{A}YL_td\mathbb{P}~\forall
A\in\mathcal{F}_t
\end{equation*}
for $\mathcal{F}_t$-measurable $Y$, under the condition that
$YL_t$ is integrable w.r.t. $\mathbb{P}$ (see
Billingsley~\cite{Billingsley95}, Theorem 16.11). In particular,
one has in terms of expectations
\begin{equation}
\tilde{E}[Y]=E[YL_t]~,\label{eq:ewrtpt}
\end{equation}
where $\tilde{E}[\cdot]$ is the expectation w.r.t.
$\tilde{\mathbb{P}}$.

On the other hand, the \textit{particular} expression of $L_t$
from Eq.~(\ref{eq:waldmart}) lends itself, by plugging in above
$Y:=\mathcal{L}_{\omega}(X)^t$ and cancelling out terms, to the
sought result, i.e.,
\begin{equation*}
E\left[e^{-\omega
S_t}\right]=\tilde{E}\left[\mathcal{L}_{\omega}(X)^t\right]~\forall
t\geq1~.
\end{equation*}
The final result, i.e., when $t$ is replaced by a stopping time
$\tau$, follows similarly by applying Theorem~3.2 from
Asmussen~\cite{Asmussen03}.

\begin{theorem}[Laplace Transform of $S_{\tau}$]\ \\\label{th:mgfst}
For an (a.s.) finite stopping time $\tau$, the Laplace transform
of the stopped sum $S_\tau$ from Eq.~(\ref{eq:stau}) is given by
\begin{equation}
\mathcal{L}_{\omega}(S_\tau)=\tilde{E}\left[\mathcal{L}_{\omega}(X)^{\tau}\right]~.\label{eq:mgfst}
\end{equation}
\end{theorem}

Note that this result is a manifestation of the earlier stated
motivation that `stopping times preserve martingale properties',
justifying thus our overall martingale framework to analyze the
inter-miss times $S_{\tau}$.

\proof\footnote{The proof proceeds along the same lines as
in~\cite{Asmussen03} with the main difference of working on an
extended product filtration. Theorem~\ref{th:mgfst} herein is thus
a simple extension of Theorem~3.2 from \cite{Asmussen03}, and no
significant technical contribution is accordingly claimed.} Fix
$T\geq0$ and choose $Y:=\mathcal{L}_{\omega}(X)^{\tau}1_{\{\tau\leq
T\}}$ which is $\mathcal{F}_T$ measurable. Applying the
integration rule from Eq.~(\ref{eq:ewrtpt}) and the properties of
conditional expectation we get
\begin{eqnarray*}
\tilde{E}\left[\mathcal{L}_{\omega}(X)^{\tau}1_{\{\tau\leq
T\}}\right]&
=&E\left[\mathcal{L}_{\omega}(X)^{\tau} 1_{\{\tau\leq T\}} L_T \right]\\
&=&E\left[E\left[\mathcal{L}_{\omega}(X)^{\tau}
1_{\{\tau\leq T\}} L_T \mid\mathcal{F}_{\tau}\right]\right]\\
&=&E\left[\mathcal{L}_{\omega}(X)^{\tau} 1_{\{\tau\leq T\}}
E\left[L_T\mid\mathcal{F}_{\tau}\right]\right]\\
&=&E\left[e^{-\omega S_\tau}1_{\{\tau\leq T\}}\right]~.
\end{eqnarray*}
In the last line we used the martingale property of $L_T$, i.e.,
$E\left[L_T\mid\mathcal{F}_{\tau}\right]=L_{\tau}$. From the
monotonicity of $1_{\{\tau\leq T\}}$ in $T$, the proof is complete
by applying Lebesgue's dominated convergence theorem (see
Theorem~16.4 in Billingsley~\cite{Billingsley95}). \hfill $\Box$

\medskip

The crucial aspect about Eq.~(\ref{eq:mgfst}) is that
$\mathcal{L}_{\omega}(X)$ is computed w.r.t. the original probability
measure $\mathbb{P}$. What remains to compute is $\tau$'s pmf
under the changed measure $\tilde{\mathbb{P}}$. In other words,
the computations for $\mathcal{L}_{\omega}(X)$ and the pmf of $\tau$
under $\tilde{\mathbb{P}}$ are entirely decoupled, circumventing
thus the dependencies in the stopped sum $S_{\tau}$.

To facilitate the auxiliary calculus under $\tilde{\mathbb{P}}$ we
next give the following technical result whose proof is deferred
to Appendix~\ref{ap:prfprop1}.

\begin{proposition}\label{prop:msptp}
On the new probability space
$(\Omega,\mathcal{F},\tilde{\mathbb{P}})$, the random variables
$X_t$ and $T_t$ have the following distribution functions for all
$t\geq1$ and $x\geq0$
\begin{eqnarray*}
\tilde{F}(x)&:=&\tilde{\mathbb{P}}\left(X_t\leq
x\right)=\frac{E\left[e^{-\omega X_t}1_{\{X_t\leq
x\}}\right]}{\mathcal{L}_{\omega}(X)}\\
\tilde{G}(x)&:=&\tilde{\mathbb{P}}\left(T_t\leq x\right)=G(x)~.
\end{eqnarray*}
The corresponding densities are
$\tilde{f}(x):=d\tilde{F}(x)=\frac{e^{-\omega
x}f(x)}{\mathcal{L}_{\omega}(X)}$ and
$\tilde{g}(x):=d\tilde{G}(x)=g(x)$, respectively. Moreover, $X_t$
and $T_t$ remain independent under $\tilde{\mathbb{P}}$.
\end{proposition}

\begin{table*}
\centering \bgroup \def\arraystretch{1.5} \small
\begin{tabular}{| l | l | l | l | l |} 
  \hline
  & \MMR = \MMS & \MDR & \MDS & \MMmin\\
  \hline
  $\displaystyle \mathcal{L}_{\omega}(S_\tau)$
  & $\displaystyle \frac{\lambda}{\lambda + \omega} \frac{\mu}{\mu + \omega}$
  & $\displaystyle \frac{\lambda e^{-\lambda/\mu}}{\lambda e^{-\lambda/\mu} + \omega e^{\omega/\mu}}$
  & $\displaystyle\frac{\lambda}{\lambda + \omega} e^{-\omega/\mu}$
  & $\displaystyle\frac{\lambda}{\lambda + \omega} \frac{\mu +
    \nu}{\mu + \nu + \omega}$ \\
  \hline
  $\mathbb{P}_{S_\tau}$
  & $\mathbb{P}_{X_1 + T_1}\;\;$ ($hypo(\lambda,\mu)$)
  & open
  & $\mathbb{P}_{X_1 + 1/\mu}$
  & $hypo(\lambda,\mu+\nu)$ \\
  \hline
  $ \e{S_\tau}$
  & $\frac{\lambda + \mu}{\lambda \mu}$
  & $\frac{1}{\lambda e^{-\lambda/\mu}}$
  & $\frac{\lambda+\mu}{\lambda\mu}$
  & $\frac{\lambda+\mu+\nu}{\lambda(\mu+\nu)}$\\
  \hline
  $ \pi = h_p$
  & $\frac{\lambda}{\lambda + \mu}$
  & $ 1 - e^{-\lambda/\mu}$
  & $\frac{\lambda}{\lambda + \mu}$
  & $\frac{\lambda}{\lambda + \mu + \nu}$\\
  \hline
\end{tabular}
\egroup \caption{Explicit Laplace transforms
  $\mathcal{L}_{\omega}(S_{\tau})$, distribution law $\mathbb{P}$, expectation
  of the stopped sum $S_{\tau}$, and cache occupancy for
  several caching models. Corresponding results for
  $M$-$D$-$min(\Sigma,\mathcal{R})$ are omitted as they depend on a closed
  form of $\mathbb{P}_{S_\tau}$ for \MDR which currently remains open. ($E[X_1]=1/\lambda$,
  deterministic $T=E[T_1]=E[T^\Sigma_1]=1/\mu$, $E[T^\mathcal{R}_1]=1/\nu$)}\vspace{-1.3em}\label{tb:examples}
\end{table*}

\subsection{The Laplace Transform of Inter-Miss Times}
Here we apply Theorem~\ref{th:mgfst} to derive the particular
transforms of the inter-miss times for the \GGR and \GGS caching
model. As the result for the \GGmin cache model is notationally
complex, it is stated in Appendix~\ref{ap:ggmincor}. Note that the
stopped sum $S_{\tau}$ from Eq.~(\ref{eq:stau}) corresponds to the
inter-miss time at a \GGRN, \GGSN, or \GGM cache, depending
whether the stopping time $\tau$ is defined as in
Eqs.~(\ref{eq:str})-(\ref{eq:stmin}), respectively.

\begin{corollary}[\GGRN]\footnote{This result was previously obtained by Fofack~\et~\cite{fofack2012analysis}.}\ \\ \label{lm:ggr}
Let $\tau$ as in Eq.~(\ref{eq:str}). If $\psi(\omega) :=
\e{e^{-\omega X}1_{\{X \leq T\}}}< 1$ for some $\omega>0$, then
the Laplace transform of the inter-miss time in the \GGR model is
given by
\begin{equation}\e{e^{-\omega S_\tau}} =
\frac{\mathcal{L}_{\omega}(X)-\psi(\omega)}{1-\psi(\omega)}~.\label{eq:ggr}
\end{equation}
\end{corollary}

To derive Corollary~\ref{lm:ggr}, it is sufficient to derive the
probability mass function of $\tau$ under $\pt$, which follows a
simple geometric structure. A full proof is stated in
Appendix~\ref{ap:prfcor1}.

\begin{corollary}[\GGSN]\ \\ \label{lm:ggs} Let $\tau$ as in
Eq.~(\ref{eq:sts}). Then for some $\omega>0$ the Laplace transform
of the inter-miss time in the \GGS model is given by
\begin{equation}\e{e^{-\omega\,S_\tau}} = \sum_{t\geq1} \mgf{\omega}^t
\e{\ft^{\,t-1}(T) \; - \,\ft^{\,t}(T)}\label{eq:mgfggs}
\end{equation} where $\ft^{\,t}$ is the distribution of the $t$-fold
convolution of $X$ in the tilded probability space.
\end{corollary}

Unlike the \GGR~model, the \GGS model is more tedious to analyze
due to the expression of the stopping time $\tau$ from
Eq.~(\ref{eq:sts}). In particular, to account for the sum in the
expression of $\tau$ a convolution density is required. We state
an according definition and a subsequent proof of
Corollary~\ref{lm:ggs} in Appendix~\ref{ap:prfcor2}.

\subsection{Examples}
Here we instantiate the results from Corollaries~\ref{lm:ggr},
\ref{lm:ggs}, and \ref{lm:ggmin} (the latter is given in
Section~\ref{ap:ggmincor}). In particular, Table~\ref{tb:examples}
gives explicit expressions for the Laplace transforms, distribution
laws $\mathbb{P}$, and expectations of the stopped sum $S_{\tau}$, and
also the cache occupancies for several simple caching models. Recall
that the results for the $\mathcal{R}$ model have been previously
obtained by Fofack~\et~\cite{fofack2012analysis}. In particular, the
results for \MMR and \MMS coincide as pointed out by Fofack~\et and,
more generally, it is easy to see that \GMR and \GMS coincide in terms
of the Laplace transforms of the stopped sum and the corresponding
caching metrics (due to the TTLs' memorylessness property). An
explicit distribution law for the inter-miss process in the case of
the \MDR model is open, albeit it can be obtained numerically.

We point out that these results hold for a single cache only. That
means that the attractive addition property of the distribution
law cannot be iterated (because the inter-miss process at a \MMRN,
\MMSN, and \MMmin model is not Poisson).

\section{Feedforward Cache Networks}\label{sec:trees}
In this section we prove that MAPs are remarkably suitable to
model inter-request processes in a feedforward cache network, to
the point that the associated superposition and input-output
operations are quite straightforward.

MAPs generalize Poisson processes by allowing the inter-arrival
times to be dependent and also to belong to the broad class of
phase-type (PH) distributions (to be defined later). MAPs have
been motivated in particular by the need to mitigate the modelling
restrictions imposed by the exponential distribution. From an
analytical perspective, MAPs are quite attractive not only due to
their versatility (MAPs are in fact dense in a large class of
point processes, see Asmussen and Koole~\cite{asmussen93}), but
also due to their tractability. Let us next give a common
definition of MAPs.

\begin{definition}[Markov Arrival Process (MAP)]\label{def:map}
A Markov arrival process is defined as a pair of matrices
$(\mathbf{D}_0, \mathbf{D}_1)$ with equal dimensions, or as a
joint Markov process $(J(t),N(t))$. The matrix
$Q:=\mathbf{D}_0+\mathbf{D}_1$ is the generator of a background
Markov process $J(t)$. The matrix $\mathbf{D}_0$ is non-singular
and a subintensity\footnote{A subintensity matrix $S$ is similar
to a stochastic matrix, except that rows sum to a non-positive
value; formally, $S_{ii}<0$, $S_{ij} \geq 0$ for $i \neq j$, and
$\sum_{j=1}^m S_{ij} \leq 0~\forall i \in \{1,\dots ,m\}$.}, and
contains the rates of the so-called \textit{hidden transitions}
which govern the change of $J(t)$ only. In turn, the matrix
$\mathbf{D}_1$ contains the (positive) rates of the so-called
\textit{active transitions} which govern the change of both $J(t)$
and a counting process $N(t)$, i.e., if $J(t^-)=i$ and a
transition $(i,j)$ from $\mathbf{D}_1$ occurs at time $t$, then
$J(t)=j$ and $N(t)=N(t^-)+1$.
\end{definition}

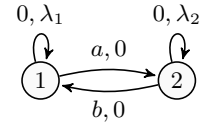
\begin{figure}[ht]
\begin{center}
\tikzstyle{every state}=[circle,fill=black!2,minimum
size=14pt,inner sep=0pt]
\begin{tikzpicture}[->,>=stealth',shorten >=1pt,auto,node distance=1.3cm,semithick]
      \node[state]                               (0) {$1$};
      \node[state,right=of 0]                    (1) {$2$};
  \draw
  (0)   edge[bend left=13] node {$a,0$}   (1)
  (0)   edge [loop above] node {$0,\lambda_1$} ()
  (1)   edge[bend left=13] node {$b,0$}   (0)
  (1)   edge [loop above] node {$0,\lambda_2$} ();;
\end{tikzpicture}
\vspace{-0.2cm} \caption{MAP representation of a MMPP; the
transitions' components are hidden and active, respectively (e.g.,
in `$0,\lambda_1$', $0$ is hidden and $\lambda_1$ is
active)}\label{fig:mmpp} \vspace{-2em}
\end{center}
\end{figure}

For the sake of familiarizing with MAPs, let us represent a
two-state Markov Modulated Poisson Process (MMPP), described in
terms of a background Markov process $J(t)$ with two states (see
Figure~\ref{fig:mmpp}); depending on the state, arrivals can occur
(and contribute to a counting process $N(t)$) at rates $\lambda_1$
and $\lambda_2$.

The corresponding MAP is given by the hidden and active transition
matrices
\begin{equation*}
\mathbf{D}_0=\left(\begin{array}{cc}-a-\lambda_1&a\\b&-b-\lambda_2\end{array}\right),~\mathbf{D}_1=\left(\begin{array}{cc}\lambda_1&0\\0&\lambda_2\end{array}\right)~,
\end{equation*}
where
$\mathbf{D}_0+\mathbf{D}_1=\left(\begin{array}{cc}-a&a\\b&-b\end{array}\right)$
is the generator of $J(t)$.

In any state, $J(t)$ is driven by two
competing transitions, i.e., a hidden one (either $a$ or $b$) and an
active one (either $\lambda_1$ or $\lambda_2$, respectively). One can
note the two emerging features of the inter-arrival times of $N(t)$:
they depend 1) on the state of the background Markov process $J(t)$,
and 2) on the time to transiting $\mathbf{D}_0$ and to make a transition
in $\mathbf{D}_1$.

\subsection{MAPs for Two Simple Cache Networks}
To introduce the main ideas of constructing MAPs for the
input-output and superposition operations, we briefly present two
simple examples of cache networks; the general results will be
presented thereafter. For further more complex examples see
Appendix~\ref{sec:fex}.

\subsubsection{Input-Output}
We first illustrate the input-output operation in a line-network
scenario as in Section~\ref{sec:lines}. Let the network from
Figure~\ref{fig:renewalcache} consist of two $\Sigma$ caching nodes
$C_1$ and $C_2$; requests arrive at $C_1$ as a Poisson process with
rate $\lambda$, and the TTLs are $\sim exp(\mu)$ and $\sim
exp(\nu)$. At node $C_1$, the arrivals can be represented as a Poisson
process $N(t)$, which is itself an elementary single-state MAP $M_1$
defined in terms of
\begin{equation*}
\mathbf{D}_0=(-\lambda),~\mathbf{D}_1=(\lambda)~, \end{equation*}
and a background Markov process with generator $\mathbf{Q}=(0)$.
See Figure~\ref{fig:renewalcache}.(a) for its graphical
representation.

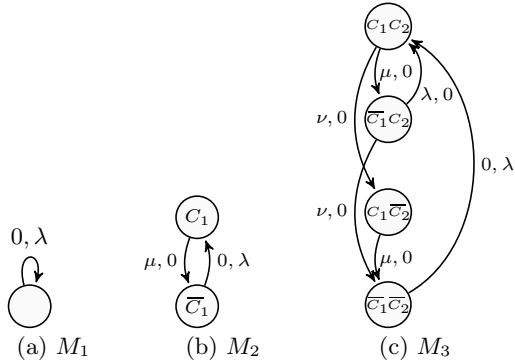
\begin{figure}[ht]
\begin{center}
\shortstack{
\begin{tikzpicture}[<-,>=stealth',shorten >=1pt,auto,node distance=1cm,semithick]
  \tikzstyle{every state}=[circle,fill=black!2,minimum size=16pt,inner sep=0pt]
\node[state]                    (2) {}; \path[->]
     (2)   edge [loop above] node {$0,\lambda$} ();
\node[draw=none,right=of 2,xshift=-0.2cm]                    (99) {};
\end{tikzpicture}\\{\hspace{-1em}\footnotesize (a) $M_1$}} \shortstack{
\begingroup
\scriptsize
\begin{tikzpicture}[->,>=stealth',shorten >=1pt,auto,node distance=0.6cm,semithick]
\tikzstyle{every state}=[circle,fill=black!2,minimum
size=16pt,inner sep=0pt]
      \node[state]                          (1) {$\overbar{C_1}$};
      \node[state, above=of 1]              (2) {$\nooverbar{C_1}$};
      \node[draw=none,right=of 2,xshift=0.4cm]                    (99) {};
   \path[->]
   (2)   edge[bend right=20] node [swap] {$\mu,0$}   (1)
   (1)   edge[bend right=20] node [swap] {$0,\lambda$}   (2);
\end{tikzpicture}
\endgroup\\{\footnotesize (b) $M_2$}}\shortstack{\begingroup
\scriptsize
\begin{tikzpicture}[->,>=stealth',shorten >=1pt,auto,node distance=0.6cm,semithick]
\tikzstyle{every state}=[circle,fill=black!2,minimum
size=16pt,inner sep=0pt]
  \node[state]                              (1) {\tiny $\overbar{C_1} \overbar{C_2}$};
  \node[state,above=of 1]                   (2) {\tiny $\nooverbar{C_1} \overbar{C_2}$};
  \node[state,above=of 2]                   (3) {\tiny $\overbar{C_1} \nooverbar{C_2}$};
  \node[state,above=of 3]                   (4) {\tiny $\nooverbar{C_1} \nooverbar{C_2}$};
  \draw
  (4)   edge[bend right=20]        node[xshift=-0.5mm] {$\mu,0$}   (3)
  (3)   edge[bend right=30]        node[swap] {$\nu,0$}    (1)
  (3)   edge[bend right=50,pos=0.4]node[swap,xshift=-1mm] {$\lambda,0$}   (4)
  (4)   edge[bend right=30]        node[swap] {$\nu,0$}   (2)
  (1)   edge[bend right=60]        node[swap] {$0,\lambda$}   (4)
  (2)   edge[bend right=20]        node[xshift=-0.5mm] {$\mu,0$}   (1);
\end{tikzpicture}
\endgroup\\{\footnotesize (c) $M_3$}
} \caption{$M_1$ corresponds to the arriving Poisson MAP at cache
  $C_1$ in Figure~\ref{fig:renewalcache}, $M_2$ to the output of
  $C_1$, and $M_3$ to the output of $C_2$}\end{center}
\vspace{-2em}
\end{figure}

To construct the arrival MAP $M_2$ at $C_2$, capturing the inter-miss
times at $C_1$, the basic idea is to duplicate the states of $M_1$ and
suitably construct the hidden and active transitions. The new states
(see Figure~\ref{fig:renewalcache}.(b)) are denoted by $\overbar{C_1}$
(with the interpretation `object is not in the cache') and $C_1$ (with
the interpretation `object is in the cache'). While in state
$\overbar{C_1}$, an arrival to $C_1$ triggers a miss---whence the
active transition $\lambda$ (i.e., the second component of
`$0,\lambda$') to $C_1$. While in state $C_1$, the TTL may expire and
hence the hidden transition $\mu$ to $\overbar{C_1}$. It is important
to remark that external requests while in $C_1$ result in hits and
thus do not affect the MAP. Note also that the constructed MAP
recovers that the inter-miss times (i.e., the time between two active
transitions) are $hypo(\lambda,\mu)$ from Table \ref{tb:examples}. In
matrix form, $M_2$ can also be represented as
\begin{equation*}
\mathbf{D}_0'=\left(\begin{array}{cc}-\lambda&0\\\mu&-\mu\end{array}\right)~\textrm{and
}\mathbf{D}_1'=\left(\begin{array}{cc}0&\lambda\\0&0\end{array}\right)~.
\end{equation*}

Applying the same idea, we construct $M_3$ by duplicating the
states of $M_2$. The four new states have the interpretations
`object is in none of the caches' (state
$\overbar{C_1}\overbar{C_2}$), `object is in only one cache'
(states $C_1\overbar{C_2}$ and $\overbar{C_1}C_2$), and `object is
in both caches' (state $C_1 C_2$). There are two important
observations to make: one is that there is a single active
transition (i.e., `$0,\lambda$') from $\overbar{C_1}\overbar{C_2}$
to $C_1C_2$. The other is that while in $\overbar{C_1}C_2$, a
request at $C_1$ does not result in an active transition because
the object is already in $C_2$---and no miss at $C_2$ can
occur---whence the \textit{hidden} transition `$\lambda,0$'. The
remaining transitions are all hidden, capturing all possible TTLs'
expirations depending on the states. In matrix form, $M''$ can
also be represented as
\begin{equation*}
\mathbf{D}_0''=\left(\begin{array}{cccc}-\lambda&0&0&0\\\mu&-\mu&0&0\\\nu&0&-\lambda-\nu&\lambda\\0&\nu&\mu&-\mu-\nu
\end{array}\right)~,
\end{equation*}
and $\mathbf{D}_1''$ contains only zeros except for $\lambda$ on
position $(1,4)$.

\subsubsection{Superposition}
\begin{figure}[ht]
\vspace{-1em}
\begin{center}
\hspace{-0.25cm}\shortstack{
\begingroup
\scriptsize \tikzstyle{every state}=[circle,fill=black!2,minimum
size=14pt,inner sep=0pt]
\begin{tikzpicture}[->,>=stealth',shorten >=1pt,auto,node distance=0.9cm,semithick]
      \node[state]                               (0) {$C_1$};
      \node[state,below=of 0]                    (1) {$\overbar{C_1}$};
   \path[->]
   (0)   edge[bend right=20] node [swap] {$\mu_1,0$}   (1)
  (1)   edge[bend right=20] node [swap] {$0,\lambda_1$}   (0);
\end{tikzpicture}
\endgroup\\{\footnotesize (a) $M_1$}}\hspace{-0.25cm} \shortstack{
\begingroup
\scriptsize \tikzstyle{every state}=[circle,fill=black!2,minimum
size=14pt,inner sep=0pt]
\begin{tikzpicture}[->,>=stealth',shorten >=1pt,auto,node distance=0.9cm,semithick]
      \node[state]                               (0) {$C_2$};
      \node[state,below=of 0]                    (1) {$\overbar{C_2}$};
   \path[->]
   (0)   edge[bend right=20] node [swap] {$\mu_2,0$}   (1)
  (1)   edge[bend right=20] node [swap] {$0,\lambda_2$}   (0);
\end{tikzpicture}
\endgroup\\{\footnotesize (b) $M_2$}}\hspace{-0.25cm}\shortstack{\begingroup
\scriptsize \tikzstyle{every state}=[circle,fill=black!2,minimum
size=14pt,inner sep=0pt]
\begin{tikzpicture}[->,>=stealth',shorten >=1pt,auto,node distance=1cm,semithick]
  \node[state]                               (00) {$C_1 C_2$};
  \node[state,right=of 00]                   (01) {$C_1 \overbar{C_2}$};
  \node[state,below=of 00]                   (10) {$\overbar{C_1} C_2$};
  \node[state,right=of 10]                   (11) {$\overbar{C_1} \overbar{C_2}$};
  \draw
  (00)   edge[bend left=7] node {$\mu_2,0$}   (01)
  (00)   edge[bend right=7] node[swap] {$\mu_1,0$}   (10)
  (01)   edge[bend left=7] node {$0,\lambda_2$}   (00)
  (10)   edge[bend right=7] node[swap] {$0,\lambda_1$}   (00)
  (01)   edge[bend left=7] node {$\mu_1,0$}   (11)
  (10)   edge[bend right=7] node[swap] {$\mu_2,0$}   (11)
  (11)   edge[bend left=7] node {$0,\lambda_1$}   (01)
  (11)   edge[bend right=7] node[swap] {$0,\lambda_2$}   (10);
\end{tikzpicture}
\endgroup\\{\footnotesize (c) $M_3$}
} \caption{The MAPs $M_1$, $M_2$, and $M_3$ corresponding to the
inter-miss times at caches $C_1$ and $C_2$ from
Figure~\ref{fig:treecaches}, and their
superposition}\label{fig:superex}
\vspace{-2em}
\end{center}
\end{figure}
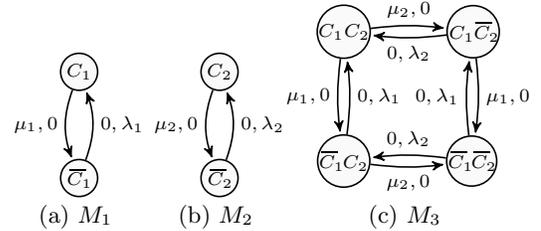

Consider now the tree topology from Figure~\ref{fig:treecaches}.
Applying the previous ideas, we can immediately construct the
(independent) MAPs $M_1$ and $M_2$ corresponding to the inter-miss
times at the caches $C_1$ and $C_2$, respectively (see
Figures~\ref{fig:superex}.(a)-(b)).

The construction of the superposition of $M_1$ and $M_2$, denoted
by $M_3$, proceeds by forming the Cartesian product of the sets of
states of $M_1$ and $M_2$; the resulting states have the same
interpretation as in the previous subsection, e.g.,
$C_1\overbar{C_2}$ stands for `object in cache $C_1$ and not in
cache $C_2$'. Moreover, the formation of the hidden and active
transitions proceeds as before. For instance, while in state
$C_1\overbar{C_2}$ two transitions are possible: an active one
(i.e., `$0,\lambda_2$') corresponding to an arrival at $C_2$, and
a hidden one (i.e., `$\mu_1,0$') corresponding to the TTL
expiration.

Furthermore, one can construct the MAP corresponding to the
inter-miss times at cache $C_3$ (in Figure~\ref{fig:treecaches})
following the ideas so far. As the resulting number of states is
eight (i.e., from doubling the states of $M_3$), we omit the
graphical depictions. We can remark however that both the
input-output and superposition operations result in an exponential
increase of the number of MAP states; this fact will be elaborated
more precisely later.

\subsection{General Results}
We now present the general results for constructing MAPs in
feedforward networks of $\mathcal{R}$, $\Sigma$, and \SR caches. As we
previously mentioned, the MAP framework allows TTLs to belong to the
broad class of PH distributions, which we define next.

\begin{definition}[Phase-Type Distribution]\label{def:ph}
Let $\mathbf{S}$ a $m \times m$ subintensity, $\mathbf{S}_0:=
-\mathbf{S} \textbf{1}$, and $\mathbf{\pi}$ a stochastic
$m$-vector. Define a Markov process with generator
  \begin{align*}
    \mathbf{P} := \begin{pmatrix}
      0 & \mathbf{0} \\
      \mathbf{S}_0 & \mathbf{S}
    \end{pmatrix}~,
  \end{align*}
which extends $\mathbf{S}$ by an absorbing state $0$ and exit
transitions from every state in $\mathbf{S}$ to $0$. A PH
distribution (of order $m$), denoted as
$T=(\mathbf{S},\mathbf{\pi})$, is defined as the time until
absorption in state $0$ of the Markov process generated by $P$,
and which starts in any of the states $\{1,\dots,m\}$ according to
$\mathbf{\pi}$.
\end{definition}
We remark that we chose the less standard notation with the
$\mathbf{0}$ vector on the first row instead of the last; this
choice will permit expressing the input-output cache operation in
a convenient manner.

Next we summarize the known result of MAPs' superposition and then
present our main results on the input-output cache operation
involving MAP requests and PH TTLs.

\subsubsection{Superposition}
First we briefly review the superposition of MAPs, for which  we
need to introduce the Kronecker sum $\oplus$ and product $\otimes$
operators for matrices.

If $\mathbf{A}$ and $\mathbf{B}$ are $m\times m$ and $n\times n$
matrices then
\begin{eqnarray*}
\mathbf{A}\otimes
\mathbf{B}&:=&\left(\begin{array}{ccc}a_{11}\mathbf{B}&\cdots&a_{1m}\mathbf{B}\\\vdots&\ddots&\vdots\\a_{n1}\mathbf{B}&\cdots&a_{mm}\mathbf{B}\end{array}\right)\\\mathbf{A}\oplus
\mathbf{B}&:=&\mathbf{A}\otimes\mathbf{I}_n+\mathbf{I}_m\otimes
\mathbf{B}~,
\end{eqnarray*}
where $\mathbf{A}=\left(a_{i,j}\right)$ and $\mathbf{I}_k$ is the
$k\times k$ identity matrix (note: the operator $\oplus$ is simplified
for the case of square matrices).

\begin{theorem}[MAP Superposition~\cite{lucantoni1990single}]
\label{thm:superpos} If the MAPs $M_1, \dots, M_n$ are represented
in terms of the matrices $(\mathbf{D}_0^1,\mathbf{D}_1^1),$\\
$\dots,(\mathbf{D}_0^n,\mathbf{D}_1^n)$, then their superposition
$M$ is also a MAP given by
\begin{eqnarray*}\mathbf{D}_0 &=& \mathbf{D}_0^1 \oplus \dots \oplus
\mathbf{D}_0^n\\\mathbf{D}_1 &=& \mathbf{D}_1^1 \oplus \dots
\oplus \mathbf{D}_1^n~.\end{eqnarray*}
\end{theorem}
With abuse of notation we use the same operator $\oplus$ for the
MAPs' superposition, i.e.,
$$M=M_1\oplus\dots\oplus M_n~.$$

Consider for example $M_3=M_1\oplus M_2$ for the MAPs from
Figure~\ref{fig:superex}, and in particular the corresponding
matrices of hidden transitions
\begin{equation*}
D_0^1=\left(\begin{array}{cc}-\lambda_1&0\\\mu_1&-\mu_1\end{array}\right),~D_0^2=\left(\begin{array}{cc}-\lambda_2&0\\\mu_2&-\mu_2\end{array}\right)~.
\end{equation*}
Then the Kronecker sum $D_0^1\oplus D_0^2$ can be written as
\begin{equation*}
\left(\begin{array}{cccc}-\lambda_1&0&0&0\\0&-\lambda_1&0&0\\\mu_1&0&-\mu_1&0\\0&\mu_1&0&-\mu_1\end{array}\right)+\left(\begin{array}{cccc}-\lambda_2&0&0&0\\\mu_2&-\mu_2&0&0\\0&0&-\lambda_2&0\\0&0&\mu_2&-\mu_2\end{array}\right)~.
\end{equation*}
It is instructive to observe that the state-space of the Kronecker
sum corresponds to the Cartesian product of the state spaces of
$M_1$ and $M_2$ in lexicographical order, and which retains the
Markovian properties of the (independent) superposed MAPs. Every
state in $M_1$ corresponds to a block of $2$ (i.e., the
dimensionality of $M_2$) states in $M_1\oplus M_2$; moreover,
every state within such a block corresponds to a state in $M_2$.
For the MAP $M_3$ from Figure~\ref{fig:superex}.(c), the
corresponding states are, in order, $\overbar{C_1}\overbar{C_2}$,
$\overbar{C_1}C_2$, $C_1\overbar{C_2}$, and $C_1C_2$.

\subsubsection{Input-Output: $\Sigma$, $\mathcal{R}$, and $\min(\Sigma,\mathcal{R})$ Caches}
Let $M$ be the MAP of cache requests and $T$ be the TTL's PH
distribution. We now prove that the input/miss process $M'$,
denoted formally using the notation
\begin{equation*}
M':=M\oslash T~,
\end{equation*}
is also a MAP, for all $\Sigma$, $\mathcal{R}$, and \SR caches.
Note that, unlike in Section~\ref{sec:lines} where the
$\mathcal{R}$ model was simpler than the $\Sigma$ model, the
opposite holds for MAPs for which reason we start with $\Sigma$.

\begin{theorem}[$MAP$-$PH$-$\Sigma$ Cache]
\label{thm:pmaphs} Consider a $\Sigma$-cache where requests arrive
according to a MAP $M=(\mathbf{D}_0,\mathbf{D}_1)$. The TTLs are
iid with a PH-distribution $T$ and generator $\mathbf{P}$; also,
$M$ and $T$ are independent. Then $M' := M \oslash T$ is a MAP
with
\begin{eqnarray*}
  \mathbf{D}_0'&=& (\mathbf{P} \oplus \mathbf{D}_0) +
  \begin{pmatrix}
    \mathbf{0} & \mathbf{0}   & \dots&\mathbf{0}\\
    \mathbf{0} & \mathbf{D}_1 & \ddots&\vdots\\
    \vdots & & \ddots &\mathbf{0}\\
    \mathbf{0} & \dots & \mathbf{0}   & \mathbf{D}_1\\
  \end{pmatrix}\\
  \mathbf{D}_1'&=&
  \begin{pmatrix}
    \mathbf{0} & \pi_1 \mathbf{D}_1 & \pi_2 \mathbf{D}_1 &\dots & \pi_m \mathbf{D}_1 \\
    \mathbf{0} & \mathbf{0} & \mathbf{0} & \dots & \mathbf{0} \\
    \vdots & \vdots  & \vdots & & \vdots\\
    \mathbf{0} & \mathbf{0} & \mathbf{0} & \dots &\mathbf{0}
  \end{pmatrix}~,
\end{eqnarray*}
\end{theorem}
where the $\mathbf{0}$ vectors have dimension $n\times n$. If $M$ has $n$
states and $T$ has $m$ transient and one absorbing states, then
$\mathbf{D}_0'$ and $\mathbf{D}_1'$ are $n(m+1) \times n(m+1)$
matrices.

\medskip

The proof is based on a constructive argument.

\proof First, it is easy to check that $M'$ is a MAP according to
Definition~\ref{def:map}. The state space of $M'$ is the Cartesian
product of the state spaces of $T$ and $M$ (thus the term
$\mathbf{P} \oplus \mathbf{D}_0$ in the expression of
$\mathbf{D}_0'$). Note that, for technical reasons, the
order of $T$ and $M$ in the Cartesian product is the opposite to
the order in $M\oslash T$. The Cartesian product accounts for all
the combinations of states from $T$ and $M$. In particular, every
state in $T$ corresponds to a block of $n$ states in $M'$ (e.g.,
the first $n$ rows and columns in $\mathbf{D}_0'$ and
$\mathbf{D}_1'$), each corresponding to a state in $M$ (recall the
example after Theorem~\ref{thm:superpos}); moreover, block $i$
corresponds to the states $(i-1)n+j~\forall j=1,\dots,n$.

Next, to prove that $M'$ models the miss process, we divide
the $m+1$ blocks of $M'$ into two groups: $\mathsf{OUT}$ and
$\mathsf{IN}$. The $\mathsf{OUT}$ group accounts for the situation
when the object is `out of the cache' and corresponds to the
absorbing state of $T$, i.e., when the TTL is expired. While in
any of the $\mathsf{OUT}$ states (corresponding to a position
$(i,j)$ in $\mathbf{D}_0'$ and $\mathbf{D}_1'$ with $1\leq i\leq
n$ and $1\leq j\leq n(m+1)$), there are both hidden transitions
(only due to the second Kronecker product in $\mathbf{P} \oplus
\mathbf{D}_0$; the first product does not contribute because the
current state of $T$ is absorbing according to our representation
of $\mathbf{P}$ from Definition~\ref{def:ph}) and active
transitions (see the first block of rows in $\mathbf{D}_1'$).  An
active transition regenerates the phase of the TTL
according to the stationary distribution $\mathbf{\pi}$ and
consequently $M'$ jumps to an $\mathsf{IN}$ block.

The $\mathsf{IN}$ group accounts for the situation when the object
is `in the cache', and each block within corresponds to one of the
phases of $T$. While in any of the $\mathsf{IN}$ states
(corresponding to a position $(i,j)$ in $\mathbf{D}_0'$ and
$\mathbf{D}_1'$ with $(n+1)\leq i\leq n(m+1)$ and $1\leq j\leq
n(m+1)$) there are only hidden transitions. Some are given by the
entries of $\mathbf{P} \oplus \mathbf{D}_0$, and thus modelling
the joint evolution of $M$ and $T$. Importantly, we remark that
since $M'$ is within an $\mathsf{IN}$ group, the active
transitions from $\mathbf{D}_1$ become passive; this is expressed
in the second term of $\mathbf{D}_0'$. Moreover, the time between
any two consecutive such transformed passive transitions
corresponds to an element $X_s$ from the definition of the
stopping time of a $\Sigma$-cache (recall Eq.~(\ref{eq:sts})).
Finally, $M'$ eventually jumps to the $\mathsf{OUT}$ block when an
exit transition from $\mathbf{T}$ occurs.

Note that the proof implicitly uses the fact that the
superposition of independent MAPs retains the underlying Markovian
properties.\hfill $\Box$

\medskip

Constructing the output for a $\mathcal{R}$ cache follows along
the same lines except that the state of the TTL is reset with each
arrival while the object is in the cache. This difference is
modelled explicitly in the second term of $D'_0$ in the following
theorem.

\begin{theorem}[$MAP$-$PH$-$\mathcal{R}$ Cache]
\label{thm:pmaphr} Under the same conditions as in
Theorem~\ref{thm:pmaphs}, but for a $\mathcal{R}$ cache, $M' := M
\oslash T$ is a MAP with
\begin{eqnarray*}
  \mathbf{D}_0'&=& (\mathbf{P} \oplus \mathbf{D}_0) +
  \begin{pmatrix}
    \mathbf{0} & \mathbf{0}          & \mathbf{0}         & \dots & \mathbf{0}\\
    \mathbf{0} & \pi_1 \mathbf{D}_1   & \pi_2 \mathbf{D}_1 & \dots & \pi_{m} \mathbf{D}_1 \\
    \vdots & \vdots & \vdots   & \dots & \vdots & \\
    \mathbf{0} & \pi_1 \mathbf{D}_1   & \pi_2 \mathbf{D}_1 & \dots & \pi_{m} \mathbf{D}_1 \\
  \end{pmatrix}\\
  \mathbf{D}_1'&=&
  \begin{pmatrix}
    \mathbf{0}& \pi_1 \mathbf{D}_1 & \pi_2 \mathbf{D}_1 &\dots & \pi_m \mathbf{D}_1 \\
    \mathbf{0} & \mathbf{0} & \mathbf{0} & \dots & \mathbf{0} \\
    \vdots & \vdots  & \vdots & & \vdots\\
    \mathbf{0} & \mathbf{0} & \mathbf{0} & \dots & \mathbf{0}
  \end{pmatrix}~.
\end{eqnarray*}
where the $\mathbf{0}$ vectors have dimension $n\times n$. If $M$ has $n$
states and $T$ has $m$ transient and one absorbing states, then
$\mathbf{D}_0'$ and $\mathbf{D}_1'$ are $n(m+1) \times n(m+1)$
matrices.
\end{theorem}

\proof The proof is identical to the previous one, except for
accounting for the difference between $\Sigma$ and $\mathcal{R}$
caches (see Eq.~(\ref{eq:sts}) vs. Eq.~(\ref{eq:str})).
Concretely, while in the states of the $\mathsf{IN}$ group, an
active transition becomes passive (as in the $\Sigma$ case), but
it also resets the phase of the TTL according to the probability vector
$\mathbf{\pi}$. \hfill $\Box$

\medskip

Finally, the case of a $\min(\Sigma,\mathcal{R})$ cache exploits
the known property that PH-distributions are closed under the
minimum operator~\cite{breuer2005introduction}):

\begin{lemma}[Minimum of two PH-distributions]\label{lma:minph}
Let $T_1 = (S_1,\pi_1)$ of order $m$, and $T_2 = (S_2,\pi_2)$ of
order $q$ be two PH distributions. Then $\min(T_1,T_2)$ is a PH
distribution of order $mq$, and given by $(S,\pi)$ where
\begin{equation*}
  S=S_1 \oplus S_2 \textrm{ and } \pi=\pi_1 \otimes \pi_2~.
\end{equation*}
\end{lemma}

The construction of the $\min(\Sigma,\mathcal{R})$ output next
follows by leveraging the property that the minimum of the two
stopping times corresponding to $\Sigma$ and $\mathcal{R}$,
respectively, carries over to the TTLs' PH representations. In
fact, given the minimum PH distribution from
Lemma~\ref{lma:minph}, the construction of the new matrix $D'_0$
is comparable to the one from Theorem~\ref{thm:pmaphr}, and
follows by repeating its
construction for the second term of $D'_0$.

\begin{theorem}[$MAP$-$PH$-$\min(\Sigma,\mathcal{R})$ Cache]
\label{thm:pmaphmin} Consider a $\min(\Sigma,\mathcal{R})$ cache
where requests arrive according to a MAP
$M=(\mathbf{D}_0,\mathbf{D}_1)$. The two TTLs are iid with PH
distributions $T^\Sigma$ and $T^{\mathcal{R}}$ for $\Sigma$ and
$\mathcal{R}$, respectively. Arrivals and both TTLs are
independent. If $\mathbf{P}$ denotes the generator of the PH
distribution $\min(T^\Sigma,T^{\mathcal{R}})$ with corresponding
initial vector $\pi=\pi^{\Sigma}\otimes\pi^{\mathcal{R}}$ (see
Lemma~\ref{lma:minph}, where $\pi^{\Sigma}$ and
$\pi^{\mathcal{R}}$ are the initial vectors of $T^{\Sigma}$ and
$T^{\mathcal{R}}$, respectively), then $M' := M \oslash
\min(T^\Sigma,T^{\mathcal{R}})$ is a MAP with
\begin{eqnarray*}
  \mathbf{D}_0'&=& (\mathbf{P} \oplus \mathbf{D}_0) + \\
  &&\begin{pmatrix}
    \mathbf{0}_{n\times n} & \mathbf{0}_{n\times nq}      &  \mathbf{0}_{n\times nq}      & \dots & \mathbf{0}_{n\times nq} \\
    \mathbf{0}_{nq\times n} & \mathbf{\Omega} &  \mathbf{0}_{nq\times
      nq}      & \dots & \mathbf{0}_{nq\times nq}      \\
    & \mathbf{0}_{nq\times nq}      &  \mathbf{\Omega} & & \vdots      \\
    \vdots & \vdots      &  \ddots      & \ddots& \ \mathbf{0}_{nq\times nq}      \\
    \mathbf{0}_{nq\times n} & \mathbf{0}_{nq\times nq}      &        & \dots &  \mathbf{\Omega} \\
  \end{pmatrix}\\
  \mathbf{D}_1'&=&
  \begin{pmatrix}
    \mathbf{0}_{n\times n}& \pi_1 \mathbf{D}_1 & \pi_2 \mathbf{D}_1 &\dots & \pi_{mq} \mathbf{D}_1 \\
    \mathbf{0}_{n\times n} & \mathbf{0}_{n\times n} & \dots & \dots & \mathbf{0}_{n\times n} \\
    \vdots & \vdots  &  & & \vdots\\
    \mathbf{0}_{n\times n} & \mathbf{0}_{n\times n} & \dots & \dots & \mathbf{0}_{n\times
    n}
  \end{pmatrix}~,
\end{eqnarray*}
where
\begin{eqnarray*}
  \mathbf{\Omega}&=&
  \begin{pmatrix}
    \mathbf{\pi^{\mathcal{R}}}_1 \mathbf{D}_1 & \dots & \mathbf{\pi^{\mathcal{R}}}_q \mathbf{D}_1 \\
          \vdots                   &       &       \vdots                   \\
    \mathbf{\pi^{\mathcal{R}}}_1 \mathbf{D}_1 & \dots & \mathbf{\pi^{\mathcal{R}}}_q \mathbf{D}_1 \\
  \end{pmatrix}
\end{eqnarray*}
has dimension $(nq\times nq)$, whereas the $\mathbf{0}_{a\times
b}$ vectors have dimension $a \times b$. If $M$ has $n$ states,
$T^\Sigma$ has $m$ transient states, and $T^{\mathcal{R}}$ has $q$
transient states, then $\mathbf{D}_0'$ and $\mathbf{D}_1'$ are
$n(m\,q+1) \times n(m\,q+1)$ matrices.
\end{theorem}

We make the important observation that because the order of $T_1$
and $T_2$ from Lemma~\ref{lma:minph} matters for the order of the
states in the Markov chain of the corresponding minimum, the order
of $\min(T^\Sigma,T^{\mathcal{R}})$ cannot be interchanged without
changing the structure of $D'_0$.

\proof The eviction event for the minimum of the two stopping times as
defined in Eq.~(\ref{eq:stmin}) translates into either reaching the
accepting state of $T^\Sigma$, or reaching the accepting state of
$T^\mathcal{R}$ without an intermittent arrival. The minimum
distribution of $T^\Sigma$ and $T^{\mathcal{R}}$ captures this
behavior up to the resetting of $T^{\mathcal{R}}$ upon arrivals. As
mentioned in the proof of Theorem~\ref{thm:pmaphr}, an arrival resets
$T^\mathcal{R}$ back to its initial state defined by its initial
vector $\pi^{\mathcal{R}}$; however, the state of $T^\Sigma$ is
preserved. By Lemma~\ref{lma:minph} (and the underlying Kronecker
sum), the states in $\min(\Sigma,\mathcal{R})$ are lexicographically
ordered with the states of $\Sigma$ followed by the states of
$\mathcal{R}$.  According to this order and because the reset behavior
only changes the state of $T^\mathcal{R}$, an arrivals' effect remains
local to each diagonal block $\Omega$. Each $\Omega$ corresponds to
the $\mathcal{R}$ reset matrix, as defined in the second term of
$D'_0$ in Theorem~\ref{thm:pmaphr}.\hfill $\Box$

In order to analyze feedforward cache networks we next complete the
description of network cache operations by introducing the splitting
concept of MAPs.

\subsubsection{Probabilistic Splitting of Arrivals}

\medskip

\begin{minipage}{0.15\textwidth}
\begin{tikzpicture}[>=stealth',shorten >=1pt,auto,node distance=0.8cm,
                    semithick]
  \tikzstyle{every state}=[rectangle,fill=black!2,minimum size=11pt,inner sep=0pt]
\node[draw=none]                           (1)   {};
\node[state,above=of 1]                    (2)   {};
\node[state,above=of 2,xshift=-0.5cm]                    (3)   {};
\node[draw=none,above=of 2,xshift=0cm]                    (4)
{\dots}; \node[state,above=of 2,xshift=0.4cm] (5)   {};
\node[draw=none,above=of 2,yshift=-0.75cm] (6)
{[$p_1$\dots$p_n$]}; \draw[->]
     (1)   edge node {$M$}   (2)
     (2)   edge[pos=0.55] node[xshift=0.1cm] {}   (3)
     (2)   edge[swap,pos=0.55] node[xshift=-0.1cm] {}   (5);
\end{tikzpicture}
\vspace{-2em}

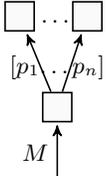
\captionof{figure}{\\n-fold split\\ of $M$} \label{fig:thin}
\end{minipage}
\begin{minipage}{0.3\textwidth}
Apart from the superposition and the input-output operations, the
third operation allows to split the MAP of an input (output)
process, as shown in Figure~\ref{fig:thin}. This works as
  an inverse operation of the superposition operator and allows to
  model the behavior of a cache feedforward network. A common
  splitting operation is when the process is split accordingly to some
  fixed probabilities. Such a construc-
\end{minipage}
tion allows to capture the behavior of an idealized load balancer.

\begin{lemma}[Splitting]
\label{lma:thin} Assume a MAP $M=(D_0,D_1)$ is split into $n$ sub
processes according to a stochastic n-vector $p$. The resulting
processes are characterized by the MAPs $M_i=(D_0^i,D_1^i)$, where
$$D_0^i=D_0+(1-p_i)D_1~\textrm{and}~D_1^i=p_i D_1~,$$ for $1\leq i \leq n$.
\end{lemma}
\proof This is an extension of the known result that a single MAP
is closed under thinning~\cite{nielsen1998note}. \hfill $\Box$

We point out that an input process represented as a MAP can be
split in different ways of which many can be captured by a
thinning operation. As further examples, the MAP arrival model is
also closed under splitting requests according to their origin, or
more generally, when the splitting decisions can be described by a
Markov process. Besides accounting for splitting operations, our
results can be further extended to account for various cache
replication strategies as considered in
Martina~\et~\cite{martina13}.

\medskip

Having introduced the main operations, we next state how to obtain
the cache metrics from a MAP model of a cache.

\subsection{Cache Metrics}

For the MAP representation, the metrics can be derived in a
uniform manner as the caching policies are encoded in the model.

\begin{lemma}[Cache Metrics for a MAP cache]
  For the \RN, \SN, and \SR caching policies, if the input process  is an
  $n$-state MAP $M=(D_0,D_1)$, with steady-state probability vector
  $p$, and the output process is an $n'$-state MAP $M'=(D_0',D_1')$, with
  steady-state vector $p'$, the miss probability and the cache occupancy are given by
  \begin{align*}
    M & = \frac{p' D_1' \mathbf{1}'}{p D_1 \mathbf{1}} \text{
      and } H = 1 - M \\
    \pi & = \sum_{i=1}^{n'} p_i \, 1_{\{\textrm{state i }\in \, \mathsf{IN}\}}
  \end{align*}
  where $\mathbf{1}$ and $\mathbf{1'}$ are all-ones vectors of dimensions $n
  \times 1$ and $n' \times 1$, respectively, and $\mathsf{IN}$ was
  defined in the proof of Theorem~\ref{thm:pmaphs}.
\end{lemma}
We note that the underlying stationary distribution of $M\oslash
T$ exists when both $M$ and $T$ are ergodic.

\subsection{Numerical Complexity} \label{sec:propcmaps} Although MAPs exactly characterize the
miss-process of a cache network, the key drawback of the
superposition and the input-output operations is that the
state-spaces of the involved MAPs increase multiplicatively in the
number of caches and the number of states of the TTLs' PH
distribution, respectively (cf.
Theorems~\ref{thm:superpos}-\ref{thm:pmaphmin}).

We summarize the scaling behavior in the following lemma for the
analysis of a binary tree.

\begin{lemma}[Scaling of State Space]
  Assume a complete binary tree of height $h$ with $2^{h-1}$ arriving
$n$-state MAPs. All nodes implement either \R or
  \S caches, with an $m$-state PH-distribution for the TTLs.  Then, for
  a fixed object, the state space size for the exact analysis of the
  miss process scales as $n^{2^h} m^{2(2^{h} -1) }$.
\end{lemma}
The proof follows immediately by induction. We note that, in the
case of \SR caches, the state space complexity is even higher.

\medskip

Although the computational complexity in the tree height appears
to be prohibitively high, we point out that the matrices arising
from the caching operations have a very regular sparse structure
(as those involved in matrix geometric
methods~\cite{Neuts1981book}). This sometimes allows to explicitly
solve for the MAP's balance equations. Besides, matrix
implementations using sparse representations enable the analysis
of medium sized caching networks. In our own implementations, used
as sanity checks for Theorems~\ref{thm:pmaphs}-\ref{thm:pmaphmin},
we considered trees with height three (without relying on
sparse-matrix representations) and trees up to height five (using
an open-source sparse-matrix implementation), with multiple MAP
arrivals and PH distributions.


\section{Conclusion}\label{sec:concl}
In this paper we have provided the first exact analysis of TTL
cache networks in great generality. We have developed two main
methods covering three common TTL caching policies: \RN, \SN, and
\SR which are employed in practical implementations. With the
first method we have generalized an existing result available for
lines of \R caches with renewal requests, by additionally
accounting for \S and \SRN. The key idea was to conveniently
formalize the three TTL caching policies by a stopped-sum
representation, whose transform could thereafter be computed using
a change of measure technique. To address the lack of closure of
renewals under superposition (and hence the inherent limitation to
line networks), our second method proposed to use the versatile
class of MAPs to model cache non-renewal requests. The key
contribution was to show that MAPs are closed under the
input-output operation of all three caching policies, whereby TTLs
follow PH distributions. This property was instrumental for the
exact analysis of feedforward networks. While the method
addressing MAPs has a much broader applicability, it suffers
however from an exponential increase in the space complexity. An
immediate research direction concerns the development of fast
numerical algorithms leveraging the sparse and regular structure
of the underlying MAPs. In this way, one could gain analytical
insight into the general cache placement problem in large
heterogeneous networks, whereby different nodes can implement
different policies.

\newpage

\input{staticbib}


\appendix
\section{Proofs of Section~4}

\subsection{Proof of Proposition~\ref{prop:msptp}}\label{ap:prfprop1}
Fix $t\geq1$ and $x\geq0$. The distribution $\tilde{F}(x)$
follows immediately from the integration rule from
Eq.~(\ref{eq:ewrtpt}):
\begin{equation*}
\tilde{\mathbb{P}}\left(X_t\leq x\right)=\int\frac{e^{-\omega
X_t}}{\mathcal{L}_{\omega}(X)}1_{\{X_t\leq x\}}d\mathbb{P}_{X_t}~,
\end{equation*}
where $\mathbb{P}_{X_t}$ is the projection of $\mathbb{P}$ on
$\sigma(X_t)$. In turn, for $\tilde{T}(x)$, we have similarly
\begin{eqnarray*} \tilde{\mathbb{P}}\left(T_t\leq
x\right)&=&\int_{\Omega_{X_t}\times\{T_t\leq x\}}\frac{e^{-\omega
X_t}}{\mathcal{L}_{\omega}(X)}d\mathbb{P}_{X_t}\times\mathbb{P}_{T_t}\\
&=&\int_{T_t\leq x}\int_{\Omega_{X_t}}\frac{e^{-\omega
X_t}}{\mathcal{L}_{\omega}(X)}d\mathbb{P}_{X_t}d\mathbb{P}_{T_t}\\
&=&\int_{T_t\leq x}d\mathbb{P}_{T_t}=F_{T}(x)~,
\end{eqnarray*}
where $\Omega_{X_t}$ denotes the (projected) sample space
corresponding to $X_t$. In the first line we used the independence
of $X_t$ and $T_t$, i.e., the random vector $(X_t,T_t)$ has the
product measure $d\mathbb{P}_{X_t}\times\mathbb{P}_{T_t}$, where
$\mathbb{P}_{X_t}$ and $\mathbb{P}_{T_t}$ are the projections of
$\mathbb{P}$ on $\sigma(X_t)$ and $\sigma(T_t)$, respectively. In
the second line we used Fubini's theorem.

Lastly, consider $B_1\in\sigma(X_t)$ and $B_2\in\sigma(T_t)$.
Using again the independence of $X_t$ and $T_t$ (under
$\mathbb{P}$) and Fubini's theorem we get
\begin{eqnarray*}
&&\hspace{-0.5cm}\tilde{\mathbb{P}}\left(X_t\in B_1,T_t\in
B_2\right)=\int\frac{e^{-\omega
X_t}}{\mathcal{L}_{\omega}(X)}1_{\{X_t\in
B_1,T_t\in B_2\}}d\mathbb{P}\\
&&=\int_{\Omega_{X_t}}\frac{e^{-\omega
X_t}}{\mathcal{L}_{\omega}(X)}1_{\{X_t\in
B_1\}}d\mathbb{P}_{X_t}\int_{T_t\in
B_2}d\mathbb{P}_{T_t}\\
&&=\tilde{\mathbb{P}}\left(X_t\in
B_1\right)\tilde{\mathbb{P}}\left(T_t\in B_2\right)~,
\end{eqnarray*}
which completes the proof.\hfill $\Box$

\subsection{Proof of Corollary~\ref{lm:ggr}}\label{ap:prfcor1}

Using the integration rule from Eq.~(\ref{eq:ewrtpt}) we
first compute
\begin{equation*}
\pt(X\leq T)  = \ee{1_{\{X\leq T\}}} = \frac{\e{1_{\{X\leq T\}}
e^{-\omega X}}}{\e{e^{-\omega X}}} =
\frac{\psi(\omega)}{\mathcal{L}_{\omega}(X)}~,
\end{equation*}
such that the pmf of $\tau$ is
\begin{eqnarray*}
\pt(\tau=t) & =& \left(\pt(X\leq T)\right)^{t-1} \left(1-\pt(X\leq T)\right) \\
&=& (\frac{\psi(\omega)}{\mathcal{L}_{\omega}(X)})^{t-1} \,
(1-\frac{\psi(\omega)}{\mathcal{L}_{\omega}(X)})~.
\end{eqnarray*}
Finally, applying Theorem~\ref{th:mgfst} and manipulating
progression series yields
\begin{eqnarray*}
\e{e^{-\omega S_\tau}} & =& \ee{\mathcal{L}_{\omega}(X)^\tau}\\
&=&\sum_{t=1}^{\infty} \left(\mathcal{L}_{\omega}(X)\right)^t (\frac{\psi(\omega)}{\mathcal{L}_{\omega}(X)})^{t-1} \, (1-\frac{\psi(\omega)}{\mathcal{L}_{\omega}(X)}) \\
& =& \frac{\mathcal{L}_{\omega}(X)-\psi(\omega)}{1-\psi(\omega)}~,
\end{eqnarray*}
which completes the proof.\hfill $\Box$

\subsection{Proof of Corollary~\ref{lm:ggs}}\label{ap:prfcor2}
We first need to introduce the distribution
convolution of $S_t$, for all $t\geq1$, in the new space
$(\Omega,\mathcal{F},\tilde{\mathbb{P}})$. These are given for all
$x\geq0$ by  by $\tilde{F}^1(x):=\tilde{F}(x)$ as in
Proposition~\ref{prop:msptp} and then recursively for $t>1$ by the
convolutions
\begin{equation*}
\tilde{F}^t(x)=\int_0^x\tilde{F}^{t-1}(x-y)d\tilde{F}(y)~,
\end{equation*}
where $\tilde{F}^0(x)=0$ for $x<0$ and $\tilde{F}^0(x)=1$ for
$x\geq0$. Assume also the existence of the corresponding densities
$\tilde{f}^t$.

We next need the pmf of $\tau$, which here proceeds by first
conditioning on $T$, and then on $S_{t-1}=X_1+\dots+X_{t-1}$ and
recalling from Proposition~\ref{prop:msptp} that $\tilde{g}(x)=g(x)$.
\begin{eqnarray*}
\pt(\tau=t) & =& \pt(S_t > T,\, S_{t-1} \leq T) \\
& =& \int_{0}^\infty \pt(S_t > x,\, S_{t-1} \leq x) \,\tilde{g}(x) \,dx \\
& =& \int_{0}^\infty \int_{0}^{x} (1-{\tilde F}(x - y)) \,{\dft^{\,t-1}}(y) \,dy \;  g(x) \,dx \\
& =& \int_{0}^\infty \left( {\ft^{\,t-1}}(x) \; - \,{\ft^{\,t}}(x)
\right) g(x) \,dx~.
\end{eqnarray*}
The rest proceeds as in the proof of Corollary~\ref{lm:ggr}.
\hfill $\Box$

\subsection{Corollary for the \GGmin Case}\label{ap:ggmincor}

\begin{corollary}[\GGminN]\ \\ \label{lm:ggmin} Let $\tau$ be defined
  as in Eq.~(\ref{eq:stmin}) and $G(\cdot)$, $g(\cdot)$ be the
  distribution and density of $T^\Sigma$, $H(\cdot)$ be the
  distribution of $T^\mathcal{R}$, and $F(\cdot)$ be the distribution
  of the arrivals.  Then for some $\omega>0$ the Laplace transform of
  the inter-miss time in the \GGmin model is given by
\begin{align*}&\e{e^{-\omega\,S_\tau}} = \\
&\sum_{t\geq1} \mgf{\omega}^t
\int_0^\infty \ee{\prod_{i=1}^{t-1}(1-H(X_i))
    \ee{H(X_t)} | \sum_{i=1}^{t-1} X_i \leq k} \\
&   - \ee{\prod_{i=1}^{t-1}(1-H(X_i))H(X_t) |
    \sum_{i=1}^{t} X_i \leq k} g(k) dk \\
&+
\int_0^\infty \ee{\prod_{i=1}^{t-1}\left( 1-H(X_i)\right)
    \ee{H(X_t)} | \sum_{i=1}^{t-1} X_i \leq k} g(k) dk \\
&+
\int_0^\infty \ee{\prod_{i=1}^{t-1}(1-H(X_i))
    \ee{1-H(X_t)} | \sum_{i=1}^{t-1} X_i \leq k} \\
&   - \ee{\prod_{i=1}^{t}(1-H(X_i)) |
    \sum_{i=1}^{t} X_i \leq k} g(k) dk
\end{align*}
\end{corollary}

The proof is quite tedious and follows by conditioning; an
explicit result for \MMmin is given in Table~\ref{tb:examples}.

\section{Examples for Caches with PH TTLs}\label{sec:fex}

In this section we give examples for the application of
Theorems~\ref{thm:pmaphs}-\ref{thm:pmaphmin}. For all of the
following three cache models (\SN, \RN, \SRN), we assume MMPP
arrivals, denoted by $M$, in the following reproduction of
Figure~\ref{fig:mmpp}:

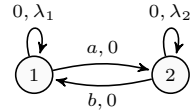
\begin{tikzpicture}[->,>=stealth',shorten >=1pt,auto,node distance=1.3cm,semithick]
\scriptsize \tikzstyle{every state}=[circle,fill=black!2,minimum
size=14pt,inner sep=0pt]
      \node[state]                               (0) {$1$};
      \node[state,right=of 0]                    (1) {$2$};
  \draw
  (0)   edge[bend left=13] node {$a,0$}   (1)
  (0)   edge [loop above] node {$0,\lambda_1$} ()
  (1)   edge[bend left=13] node {$b,0$}   (0)
  (1)   edge [loop above] node {$0,\lambda_2$} ();;
\end{tikzpicture}

\subsection{$\Sigma$ Cache Model}

We assume the following TTL $T$ (in Markov chain representation
and which starts in state $1$ with probability one).

\begin{tikzpicture}[->,>=stealth',shorten >=1pt,auto,node
  distance=1.3cm,semithick]
\scriptsize \tikzstyle{every state}=[circle,fill=black!2,minimum
size=14pt,inner sep=0pt]
      \node[state]                               (1) {$1$};
      \node[state,right=of 1]                               (2) {$2$};
      \node[state,accepting,right=of 2]          (0) {$0$};
  \draw
  (1)   edge[] node {$\mu_1$}   (2)
  (2)   edge[dashed] node {$\mu_2$}   (0);
\end{tikzpicture}

The output MAP is constructed by replicating the MMPP's states for
each state of the TTL and adjusting for the inherent cache
property, that no misses occur while the object is in the cache.
This basic idea is reflected in taking the Cartesian product of
$T$ and $M$ and subsequently making all of $D_1$'s transitions
passive (cf. definition of $D_0'$ in Theorem~\ref{thm:pmaphs}).

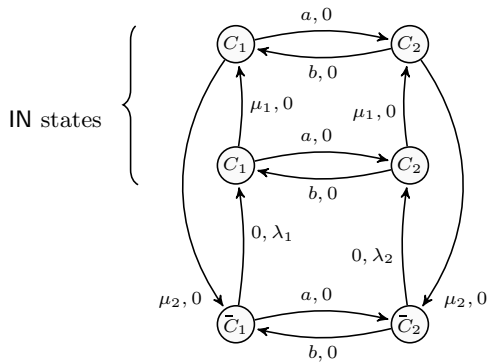
\begin{tikzpicture}[->,>=stealth',shorten >=1pt,auto,node distance=1.1cm,semithick]
\scriptsize \tikzstyle{every state}=[circle,fill=black!2,minimum
size=14pt,inner sep=0pt]
      \node[state]                               (0o) {$\overbar{C}_1$};
      \node[state,right=of 0o,xshift=0.7cm]      (1o) {$\overbar{C}_2$};
      \node[state,above=of 0o,yshift=0.5cm]      (0i1) {$\nooverbar{C}_1$};
      \node[state,above=of 1o,yshift=0.5cm]      (1i1) {$\nooverbar{C}_2$};
      \node[state,above=of 0i1]                  (0i2) {$\nooverbar{C}_1$};
      \node[state,above=of 1i1]                  (1i2) {$\nooverbar{C}_2$};
  \draw
  (0o)   edge[bend left=13,pos=0.45]      node {$a,0$}             (1o)
  (1o)   edge[bend left=13]               node {$b,0$}             (0o)
  (0i1)   edge[bend left=13,pos=0.45]     node {$a,0$}             (1i1)
  (1i1)   edge[bend left=13]              node {$b,0$}             (0i1)
  (0i2)   edge[bend left=13,pos=0.45]     node {$a,0$}             (1i2)
  (1i2)   edge[bend left=13]              node {$b,0$}             (0i2)
  (0o)   edge[bend right=8,swap,pos=0.6]  node {$0,\lambda_1$}     (0i1)
  (1o)   edge[bend left=8,pos=0.4]        node {$0,\lambda_2$}     (1i1)
  (0i1)  edge[bend right=8,swap,pos=.5]   node {$\mu_1,0$}         (0i2)
  (1i1)  edge[bend left=8,pos=0.4]        node {$\mu_1,0$}         (1i2)
  (0i2)  edge[bend right=35,swap,pos=0.9] node {$\mu_2,0$}         (0o)
  (1i2)  edge[bend left=35,pos=0.9]       node {$\mu_2,0$}         (1o);
   \draw[-,decorate,decoration={brace,amplitude=0.2cm}] ([xshift=-1.3cm]0i1.south) -- ([xshift=-1.3cm]0i2.north) node[yshift=-1.25cm,xshift=-1.1cm] {\normalsize $\mathsf{IN}$ states};
\end{tikzpicture}

Note that we do not draw self-loops unless they are active
transitions (i.e., the entries $D'_{0_{ii}}$ are not drawn,
whereas an entry $D'_{1_{ii}}$ is drawn, as in the MMPP example).
While the cache is in the $\mathsf{IN}$ state, further arrivals do
not change the state of the cache. Thus, there are no transitions
with $\lambda_1$ or $\lambda_2$ in the $\mathsf{IN}$ part of the
resulting cache.

\subsection{$\mathcal{R}$ Cache Model}

Similarly to the $\Sigma$ case we consider the MMPP arrival
process $M$ and the following TTL $T$ (in Markov chain
representation and which starts in state $1$ with probability
one):

\begin{tikzpicture}[->,>=stealth',shorten >=1pt,auto,node
  distance=1.3cm,semithick]
\scriptsize \tikzstyle{every state}=[circle,fill=black!2,minimum
size=14pt,inner sep=0pt]
      \node[state]                               (1) {$1$};
      \node[state,right=of 1]                               (2) {$2$};
      \node[state,accepting,right=of 2]          (0) {$0$};
  \draw
  (1)   edge[] node {$\nu_1$}   (2)
  (2)   edge[dashed] node {$\nu_2$}   (0);
\end{tikzpicture}

Constructing the output MAP for the $\mathcal{R}$ case bears a
subtle difference from the $\Sigma$ case.  The basic idea is again
to replicate the MMPP's states for each state of the TTL but then
we have to accommodate for the $\mathcal{R}$ resetting behavior of
this cache model: every arrival while the object is in the cache
resets the TTL's state according to its initial vector.

Recalling the notations from Theorem~\ref{thm:pmaphr}, this idea
is reflected by taking the Cartesian product of $T$ and $M$ and
subsequently adjusting for the ``resetting behavior'', i.e., by
making $D_1$'s transitions passive and resetting $T$'s state.

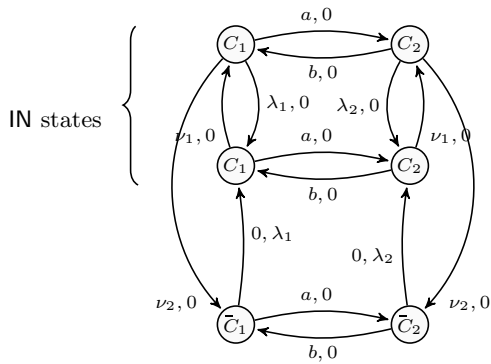
\begin{tikzpicture}[->,>=stealth',shorten >=1pt,auto,node distance=1.1cm,semithick]
\scriptsize \tikzstyle{every state}=[circle,fill=black!2,minimum
size=14pt,inner sep=0pt]
      \node[state]                               (0o) {$\overbar{C}_1$};
      \node[state,right=of 0o,xshift=0.7cm]      (1o) {$\overbar{C}_2$};
      \node[state,above=of 0o,yshift=0.5cm]      (0i1) {$\nooverbar{C}_1$};
      \node[state,above=of 1o,yshift=0.5cm]      (1i1) {$\nooverbar{C}_2$};
      \node[state,above=of 0i1]                  (0i3) {$\nooverbar{C}_1$};
      \node[state,above=of 1i1]                  (1i3) {$\nooverbar{C}_2$};
  \draw
  (0o)   edge[bend left=13,pos=0.45]      node {$a,0$}             (1o)
  (1o)   edge[bend left=13]               node {$b,0$}             (0o)
  (0i1)   edge[bend left=13,pos=0.45]     node {$a,0$}             (1i1)
  (1i1)   edge[bend left=13]              node {$b,0$}             (0i1)
  (0i3)   edge[bend left=13,pos=0.45]     node {$a,0$}             (1i3)
  (1i3)   edge[bend left=13]              node {$b,0$}             (0i3)
  (0o)   edge[bend right=8,swap,pos=0.6]  node {$0,\lambda_1$}     (0i1)
  (1o)   edge[bend left=8,pos=0.4]        node {$0,\lambda_2$}     (1i1)
  (0i3)   edge[bend left=30,pos=0.48]     node {$\lambda_1,0$}     (0i1)
  (1i3)   edge[bend right=30,swap,pos=0.52] node {$\lambda_2,0$}     (1i1)
  (0i1)  edge[bend left=18,pos=0.3]         node {$\nu_1,0$}         (0i3)
  (1i1)  edge[bend right=18,swap,pos=0.3]   node {$\nu_1,0$}         (1i3)
  (0i3)  edge[bend right=43,swap,pos=0.9]   node {$\nu_2,0$}         (0o)
  (1i3)  edge[bend left=43,pos=0.9]         node {$\nu_2,0$}         (1o);
   \draw[-,decorate,decoration={brace,amplitude=0.2cm}] ([xshift=-1.3cm]0i1.south) -- ([xshift=-1.3cm]0i3.north) node[yshift=-1.25cm,xshift=-1.1cm] {\normalsize $\mathsf{IN}$ states};
\end{tikzpicture}

Note that if there was a greater number of TTL states and a
non-trivial initial probability vector $\pi$ for $T$, then the
passive transitions `$\lambda_1,0$' and `$\lambda_2,0$' for each
state of $T$ would be directed to the initial states according to
$\pi$ and independently of $T$'s current state. This is
represented by the second term of $D_0'$ in
Theorem~\ref{thm:pmaphr} by repeating the row with $\pi_i D_1$ for
each state of the TTL.

\medskip

Finally, we turn to the \SR cache model which is more complicated
due to the higher number of states involved.

\subsection{$\min(\Sigma,\mathcal{R})$ Cache Model}

Consider the same MMPP arrival process $M$ with the following TTL
representations. For the $\Sigma$ part of the model, the TTL is
called $T^\Sigma$:

\begin{tikzpicture}[->,>=stealth',shorten >=1pt,auto,node
  distance=1.3cm,semithick]
\scriptsize \tikzstyle{every state}=[circle,fill=black!2,minimum
size=14pt,inner sep=0pt]
      \node[state]                               (1) {$1$};
      \node[state,right=of 1]                               (2) {$2$};
      \node[state,accepting,right=of 2]          (0) {$0$};
  \draw
  (1)   edge[] node {$\mu_1$}   (2)
  (2)   edge[dashed] node {$\mu_2$}   (0);
\end{tikzpicture}

In turn, for the $\mathcal{R}$ part of the model, the TTL is
called $T^\mathcal{R}$:

\begin{tikzpicture}[->,>=stealth',shorten >=1pt,auto,node
  distance=1.3cm,semithick]
\scriptsize \tikzstyle{every state}=[circle,fill=black!2,minimum
size=14pt,inner sep=0pt]
      \node[state]                               (1) {$1$};
      \node[state,right=of 1]                               (2) {$2$};
      \node[state,accepting,right=of 2]          (0) {$0$};
  \draw
  (1)   edge[] node {$\nu_1$}   (2)
  (2)   edge[dashed] node {$\nu_2$}   (0);
\end{tikzpicture}

The corresponding output for a $\min(\Sigma,\mathcal{R})$ cache
follows by constructing the PH minimum for
$\min(T^\Sigma,T^\mathcal{R})$ which has four transient states and
one absorbing state.

Then, we replicate the MMPP's states for each state of
$\min(T^\Sigma,T^\mathcal{R})$ and link this Cartesian product
construction with the reset behavior of the $\mathcal{R}$ model.
As pointed out in the proof of Theorem~\ref{thm:pmaphmin}, the
resetting behavior of $\mathcal{R}$ has to preserve the state of
$T^\Sigma$. This behavior is represented in the following output
MAP by the two `$\lambda_1,0$' and the two `$\lambda_2,0$'
transitions.

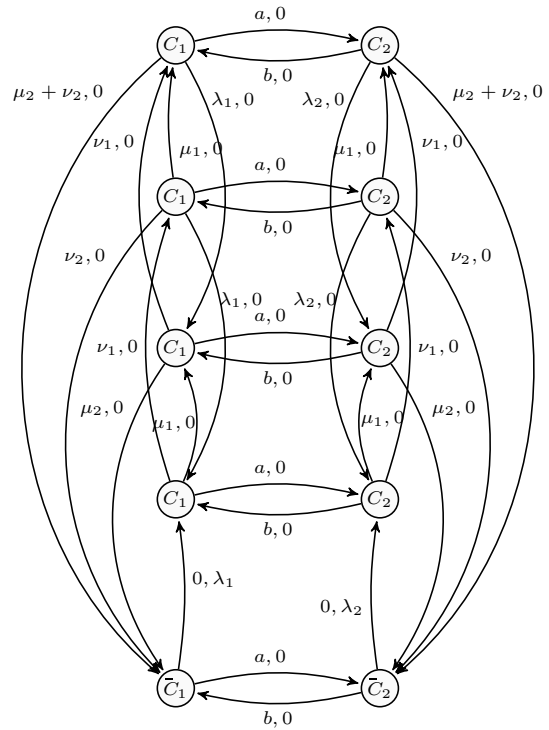
\begin{tikzpicture}[->,>=stealth',shorten >=1pt,auto,node distance=1.5cm,semithick]
\scriptsize \tikzstyle{every state}=[circle,fill=black!2,minimum
size=14pt,inner sep=0pt]
      \node[state]                               (0o) {$\overbar{C}_1$};
      \node[state,right=of 0o,xshift=0.7cm]      (1o) {$\overbar{C}_2$};
      \node[state,above=of 0o,yshift=0.5cm]      (0i1) {$\nooverbar{C}_1$};
      \node[state,above=of 1o,yshift=0.5cm]      (1i1) {$\nooverbar{C}_2$};
      \node[state,above=of 0i1]                  (S2R1M1) {$\nooverbar{C}_1$}; 
      \node[state,above=of 1i1]                  (S2R1M2) {$\nooverbar{C}_2$}; 
      \node[state,above=of S2R1M1]                  (S1R2M1) {$\nooverbar{C}_1$}; 
      \node[state,above=of S2R1M2]                  (S1R2M2) {$\nooverbar{C}_2$}; 
      \node[state,above=of S1R2M1]                  (S2R2M1) {$\nooverbar{C}_1$}; 
      \node[state,above=of S1R2M2]                  (S2R2M2) {$\nooverbar{C}_2$}; 
  \draw
  (0o)   edge[bend left=13,pos=0.45]      node {$a,0$}             (1o)
  (1o)   edge[bend left=13]               node {$b,0$}             (0o)
  (0i1)   edge[bend left=13,pos=0.45]     node {$a,0$}             (1i1)
  (1i1)   edge[bend left=13]              node {$b,0$}             (0i1)
  (S2R1M1)   edge[bend left=13,pos=0.45]     node {$a,0$}             (S2R1M2)
  (S2R1M2)   edge[bend left=13]              node {$b,0$}             (S2R1M1)
  (S1R2M1)   edge[bend left=13,pos=0.45]     node {$a,0$}             (S1R2M2)
  (S1R2M2)   edge[bend left=13]              node {$b,0$}             (S1R2M1)
  (S2R2M1)   edge[bend left=13,pos=0.45]     node {$a,0$}             (S2R2M2)
  (S2R2M2)   edge[bend left=13]              node {$b,0$}             (S2R2M1)
  (0o)   edge[bend right=8,swap,pos=0.6]  node {$0,\lambda_1$}     (0i1)
  (1o)   edge[bend left=8,pos=0.4]        node {$0,\lambda_2$}     (1i1)
  (S2R2M1)   edge[bend left=30,pos=0.2]      node[xshift=-0.1cm] {$\lambda_1,0$}     (S2R1M1)
  (S2R2M2)   edge[bend right=30,swap,pos=0.2]node[xshift=0.1cm] {$\lambda_2,0$}     (S2R1M2)
  (S1R2M1)   edge[bend left=30,pos=0.38]      node[xshift=-0.15cm] {$\lambda_1,0$}     (0i1)
  (S1R2M2)   edge[bend right=30,swap,pos=0.38]node[xshift=0.15cm] {$\lambda_2,0$}     (1i1)
  (0i1)  edge[bend right=23,pos=.5]       node[xshift=0.09cm] {$\mu_1,0$}         (S2R1M1)
  (1i1)  edge[bend left=23,swap,pos=0.4]  node[xshift=-0.09cm] {$\mu_1,0$}         (S2R1M2)
  (0i1)  edge[bend left=18]               node {$\nu_1,0$}         (S1R2M1)
  (1i1)  edge[bend right=18,swap]         node {$\nu_1,0$}         (S1R2M2)
  (S2R1M1)  edge[bend left=22,pos=0.65]       node {$\nu_1,0$}         (S2R2M1)
  (S2R1M2)  edge[bend right=22,swap,pos=0.65] node {$\nu_1,0$}         (S2R2M2)
  (S1R2M1)  edge[bend left=8,swap,pos=0.1]   node {$\mu_1,0$}         (S2R2M1)
  (S1R2M2)  edge[bend right=8,pos=0.1]       node {$\mu_1,0$}         (S2R2M2)
  (S2R1M1)  edge[bend right=35,swap,pos=0.2] node {$\mu_2,0$}         (0o)
  (S2R1M2)  edge[bend left=35,pos=0.2]       node {$\mu_2,0$}         (1o)
  (S1R2M1)  edge[bend right=45,swap,pos=0.15]node {$\nu_2,0$}         (0o)
  (S1R2M2)  edge[bend left=45,pos=0.15]      node {$\nu_2,0$}         (1o)
  (S2R2M1)  edge[bend right=50,swap,pos=0.1] node {$\mu_2+\nu_2,0$}   (0o)
  (S2R2M2)  edge[bend left=50,pos=0.1]       node {$\mu_2+\nu_2,0$}   (1o);
\end{tikzpicture}

For a better visualization, the five layers of the above output
MAP can be interpreted as 1) $\mathsf{OUT}$ (absorbing state), 2)
$\mathsf{IN}$ $\Sigma-\textrm{State~1}$
$\mathcal{R}-\textrm{State~1}$, 3) $\mathsf{IN}$
$\Sigma-\textrm{State~2}$ $\mathcal{R}-\textrm{State~1}$, 4)
$\mathsf{IN}$ $\Sigma-\textrm{State~1}$
$\mathcal{R}-\textrm{State~2}$, and 5) $\mathsf{IN}$
$\Sigma-\textrm{State~2}$ $\mathcal{R}-\textrm{State~2}$, from
bottom up, where $\textrm{OUT}$ and $\mathsf{IN}$ have the
meanings from the proof of Theorem~\ref{thm:pmaphs}.

\end{document}

%% file: staticbib.tex
\providecommand{\noopsort}[1]{}